\documentclass[%
 reprint,
 amsmath,amssymb,
 aps,
]{revtex4-1}

\usepackage{graphicx}
\usepackage{dcolumn}
\usepackage{bm}

\usepackage{epstopdf}

\begin{document}


\title{The Structured `Low Temperature' Phase of the Retinal Population Code}

\author{Mark L. Ioffe $^{a,b}$}
\author{Michael J. Berry II $^b$}%
 \email{mioffe@princeton.edu}

\affiliation{$^a$ Department of Physics, Princeton University, Princeton, New Jersey, 08544}
\affiliation{$^b$ Princeton Neuroscience Institute, Princeton University, Princeton, New Jersey, 08544}

\date{\today}

\begin{abstract}
Recent advances in experimental techniques have allowed the simultaneous recording of populations of hundreds of neurons, allowing more comprehensive investigation into the nature of the collective structure of population neural activity. While recent studies have reported a phase transition in the parameter space of maximum entropy models describing the neural probability distributions, the interpretation of these findings has been debated. Here, we suggest that this phase transition may be evidence of a `structured', collective state in the neural population. We show that this phase transition is robust to changes in stimulus ensemble and adaptive state. We find that the pattern of pairwise correlations between neurons has a strength that is well within the strongly correlated regime and does not require fine tuning, suggesting that this state is generic for populations of 100+ neurons. We find a clear correspondence between the emergence of a phase transition, and the emergence of attractor-like structure in the inferred energy landscape. A collective state in the neural population, in which neural activity patterns naturally form clusters, provides a consistent interpretation for our results. 
\end{abstract}

\maketitle

The past decade has witnessed a rapid development of new techniques for recording simultaneously from large populations of neurons \cite{marre, shlens2009, ahrens, tolias}. As the experimentally accessible populations increase in size, a natural question arises: how can we model and understand the activity of large populations of neurons? In statistical physics, the interactions of large numbers of particles create new, {\it statistical}, laws that are qualitatively different from the original mechanical laws describing individual particle interactions. Studies of these statistical laws have allowed the prediction of macroscopic properties of a physical system from knowledge of the microscopic properties of its individual particles \cite{LLv5}.  By exploiting analogies to statistical physics, one might hope to arrive at new insights about the collective properties of neural populations that are also qualitatively different from our understanding of single neurons.

The correlated nature of retinal ganglion cell spike trains can profoundly influence the structure of the neural code. Information can be either reduced or enhanced by correlations depending on the nature of the distribution of firing rates \cite{shamirsomp}, the tuning properties of individual neurons \cite{morenobote}, stimulus correlations and neuronal reliability \cite{tkacik2010,ravafranke}, the patterns of correlations \cite{ravamjb}, and interaction among all these factors \cite{wilkeeurich}. In addition, the structure of the decoding rule needed to read out the information represented by a neural population can be strongly influenced by the pattern of correlation regardless of whether it reduces or enhances the total information  \cite{averbecklathampouget,schwartzberry2012}. One approach to understanding the properties of measured neural activity is to study the nature of minimally structured (`maximum entropy') models of the probability distribution that reproduce the measured correlational structure \cite{jaynesmaxent, schneidman2006, tkacik2014}. These models have been shown to be highly accurate in reproducing the full statistics of the activity patterns of small numbers of neurons \cite{schneidman2006, shlens2006, ohiorhenuan}. The hope is that even if these models underestimate the real structure of larger neural populations, the properties of the distribution which arise in these simplified models are general and of consequence to the true distribution. 

Maximum entropy models that constrain only the pairwise correlations between neurons are generalized versions of the Ising model, one of the simplest models in statistical physics where collective effects can become significant. The macroscopic behavior of these models varies substantially depending on the parameter regime. By fitting these models to measured neural acitivity, we can begin to explore (by analogy) the `macroscopic' properties of the retinal population code. 

One macroscopic property of interest is the specific heat. Discontinuity or divergence of this property is an indicator of a phase transition, which implies a qualitative change in the properties of the system. Previous studies have shown that the specific heat has a peak that grows and sharpens as more neurons are simultaneously analyzed \cite{tkacik2015, dariothesis, mora2015, yu}. Most of the literature on this topic can be divided into two camps: the `proponents' who argue that this is a signature of criticality, i.e. the system is poised in between high and low temperature phases, in order to optimize the capacity of the neural code \cite{mora2015,tkacik2015,mora2011}, and the `sceptics' who argue that this is merely a consequence of ignored latent variables \cite{schwab, aitchison}, ignored higher order correlation structure in the data \cite{macke2011}, or even the presence of any correlation at all \cite{macke2016}. 

An alternative interpretation is that system is in a `low temperature' state. Empirically, the peak of the specific heat is found at a higher temperature than the operating point of the system ({\it T} = 1), suggesting that the system is on the low temperature side of the phase transition. Low temperature phases in statistical physics are usually associated with structure in the distribution of states, in which the system can `freeze' into ordered states. High temperature phases, in contrast, are associated with weakly correlated, nearly independent structure in the population of neurons. From this perspective, the phase transition serves as an indirect indicator of structure in the probability landscape of the neural population at the real operating point ({\it T} = 1).

Maximum entropy models fit particular statistics of the distributions of experimentally measured neural neural activity \cite{tkacik2014}. Because retinal responses are specific to both the adaptational state of the retina and the ensemble of stimuli chosen to probe them, the measured pattern of neural correlation -  and hence the detailed properties of the maximum entropy model - will also vary.  Therefore, it is yet unclear how robust is the presence of a low temperature state to different experimental conditions. The phase transition itself arises as a consequence of the correlations between neurons. This pattern of correlation in turn has contributions from correlations in the stimulus and from retinal processing. The distribution of correlations also has a particular shape, with many weak but statistically non-zero terms. It is unclear how these different properties contribute to the nature of the structured collective state of the neural population.

Here, we show that while the detailed statistics of the retinal population code differ across experimental conditions, the observed phase transition persists. We find that retinal processing provides substantial contributions to the pattern of correlations among ganglion cells and thus to the specific heat, as do the many weak but statistically non-zero correlations in the neural population. We also find that the spatio-temporal processing of the classical receptive field is not sufficient to understand the collective properties of ganglion cell populations. To address the nature of the collective state of the retinal population code, we explored how a particle representing the state of neural activity moves over the system's energy landscape under the influence of finite temperature. We find that the energy landscape has regions that ``trap'' particle motion, in analogy to basins of attraction in a dynamical system. By varying the overall correlation strength, we show that the emergence of this structure is closely connected to the emergence of the measured phase transition. This emergence occurs at surprisingly low overall correlation strength, indicating that the real population is robustly within the structured regime.

\section*{Results}

One of the main goals of our study is to test whether the collective state of a neural population is robust to different experimental conditions. Adaptation to the statistics of the visual input is a central feature of retinal processing, and any robust property of the retinal population code should be present in different adaptational states. We focused first on adaptation, and in particular we chose an experiment probing adaptation to ambient illumination.

To process the visual world, the retina has to adapt to daily variations of the ambient light level on the order of a factor of a hundred billion \cite{rodieck}. Prominent examples of known sites in the retinal circuit with adaptive mechanisms include the voltage-intensity curves of photoreceptors \cite{kleinschmidt, fain, dowlingbook}, the nonlinear output of bipolar cells \cite{grimes}, and the surround structure of ganglion cells \cite{barlow}. A significant contribution to these effects arises from the light-dependent global dopamine signal \cite{witkov1, witkov2}, which regulates retinomotor changes in the shapes of photoreceptors \cite{pierce}, and gap junction transmission in horizontal \cite{lasater} and AII amacrine cells \cite{witkov2}. The global nature of the dopamine signal suggests that any cells or synapses in the retina that possess dopaminergic receptors (almost all retinal cell types studied, \cite{nguyen-legros}), will experience adaptive effects of changes in the mean light level. Though the literature on single cell adaptation to ambient illumination is extensive, little is known about the changes in correlational structure across the population of retinal ganglion cells.

We recorded from a population of tiger salamander ({\it Ambystoma Tigrinum}) retinal ganglion cells responding to the same natural movie at two different ambient light levels. In Experiment $\#1$, we recorded with and then without an absorptive neutral density filter of optical density 3 (which attenuates the intensity of light by a factor of $10^3$) in the light path of the stimulus projecting onto the retina. The stimuli consisted of a chromatic checkerboard and a repeated natural movie (details in Methods). Thus, the contrast and statistics of each visual stimulus were the same under both luminance conditions, with the only difference being that in the {\it dark} condition, the mean light level was 1000 times lower than in the {\it light} condition.

\begin{figure*}[t]
\centerline{\includegraphics[width=17.8cm]{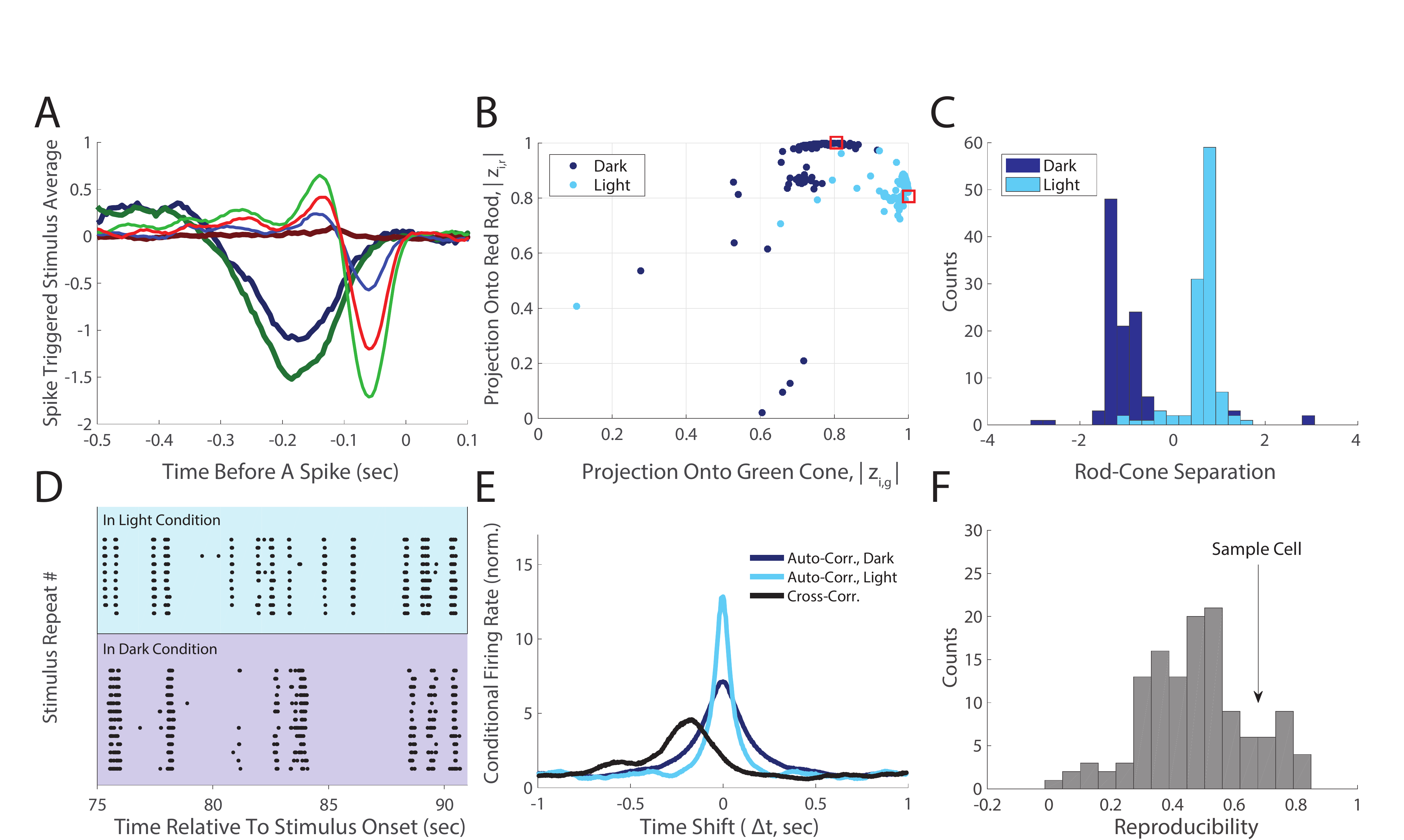}}
\caption{ {\bf Changes In Individual RGC Feature Selectivities Are Consistent With A Switch Between Rod and Cone Dominated Circuitry}. {\bf A}. Color-dependent STA for a sample ganglion cell (Experiment $\#1$). Lighter shades measured in the {\it light} condition. Red, green and blue colors correspond to the monitor gun. {\bf B}. Projections of the ganglion cell color profiles onto the red rod ($z_{i,r}$)  on the y-axis plotted versus the projection onto the green cone ($z_{i,g}$) for {\it N}=111 ganglion cells (see Supplement); dark blue dots for the {\it dark}, light blue dots for {\it light} condition. Absolute values are used to provide invariance to the sign of the color vector (ON- vs. OFF-type cells). The rod-cone overlap (similarity of the two photopigments) is plotted as red squares. {\bf C}. The distribution of the rod-cone spectral separations (see Supplement), plotted for the {\it dark} (dark blue) and {\it light} (light blue) conditions. {\bf D}. Responses of a sample ganglion cell to repeated presentations of the same stimulus segment, in the two conditions. {\bf E}. Auto- and cross-correlations of spike trains for the cell in panel {\bf D}, calculated excluding spikes from the same trial; {\it dark-dark} auto-correlation (dark blue), {\it light-light} auto-correlation (light blue), {\it dark-light} cross-correlation (black). These are normalized by the probability of a spike, so that this measure tends to 1 at long time shifts. {\bf F}. Distribution over ganglion cells of the reproducibility (see main text) of spikes across light adapted conditions, with an arrow indicating the sample cell in {\bf D,E}. }
\label{fig:fig1}
\end{figure*}

The responses of individual cells to the checkerboard allowed us to measure the spatiotemporal receptive field of each ganglion cell using reverse correlation \cite{dayanabbott, fairhall2006} in each light condition. For most cells these three linear filters were scaled versions of each other, suggesting a single filter with different sensitivities to the red, green and blue monitor guns (Fig. \ref{fig:fig1}A). The vast majority of rods in the tiger salamander retina are classified as `red rods' (98$\%$); similarly, most of the cones are `green cones' ($80\%$) \cite{sherry, perry}. We estimated the relative sensitivities of these two photopigments to our monitor guns from the reported spectra of these photopigments \cite{baylor, cornwall, perry} and a measurement of the spectral output of the monitor guns (see Supplement). We found that for many ganglion cells the relative amplitudes of these three sensitivities were closely consistent with the estimated sensitivity of the red rod photopigment in the {\it dark} recording, and the green cone photopigment in the {\it light} recording (Fig. \ref{fig:fig1}B,C). These results suggested to us that in our experiments, retinal circuitry was in the scoptopic, rod-dominated limit in our {\it dark} condition and in the photopic, cone-dominated limit in our {\it light} condition.

How much does the feature selectivity of ganglion cells change across the two light-adapted conditions? We can see from the temporal kernel of the spike-triggered average that there are some changes in temporal processing along with a big change in response latency (Fig. \ref{fig:fig1}A). Another way to compare feature selectivity is to look at what times a ganglion cell fires spikes to repeated presentations of the same natural movie (Fig. \ref{fig:fig1}D). The spike timing of individual ganglion cells was reproducible across repeats of a natural movie within a particular luminance condition. However, across conditions there were significant changes in which times during the stimulus elicited a spike from the same cell (Fig. \ref{fig:fig1}D).

To quantify this effect over our entire recording, we estimated the shuffled autocorrelation function of each cell's spike train (i.e. the correlation function between spikes on one trial and those on another trial \cite{berry1997}). The width of this shuffled autocorrelation function is one measure of timing precision for the spike train \cite{berry1997}. The narrower width of the autocorrelation curve in the {\it light} adapted condition indicated greater spike timing precision in that condition. The 200ms offset in the peak of the cross correlation curve was characteristic of the longer delays in signal processing that arise in the rod versus cone circuitry, which can also be seen in the different latencies of the reverse correlation (Fig. \ref{fig:fig1}A). The area under the curve, but above the random level set by the firing rate, is a measure of reproducibility across trials. Normalizing this measure across luminance conditions yielded an estimate of spiking reproducibility across light conditions. A value of unity for this measure indicates that spiking reproducibility across light conditions is as high as the reproducibilities within each condition. We found a wide range of reproducibility across our populations of neurons (0 to 0.75, Fig. \ref{fig:fig1}E), suggesting that significant changes in feature selectivity occured for most neurons. Our results are qualitatively consistent with a recently published study \cite{tikidji}, where retinal ganglion cells gain or lose specific firing events at different ambient light levels.

Given that our observed changes in chromatic sensitivites and feature selectivities of ganglion cells were consistent with distinct adaptational states for the retina, it seemed likely that we would find statistically significant changes in the average firing rates, pairwise correlations, and the distribution of simultaneously active cells. Because the details of the maximum entropy model are a function of this set of constraints, statistically significant changes in these moments across light conditions were particularly important. Otherwise, any observed `robustness' at the collective level would be trivial. 

\begin{figure}
\centerline{\includegraphics[width=8.6cm]{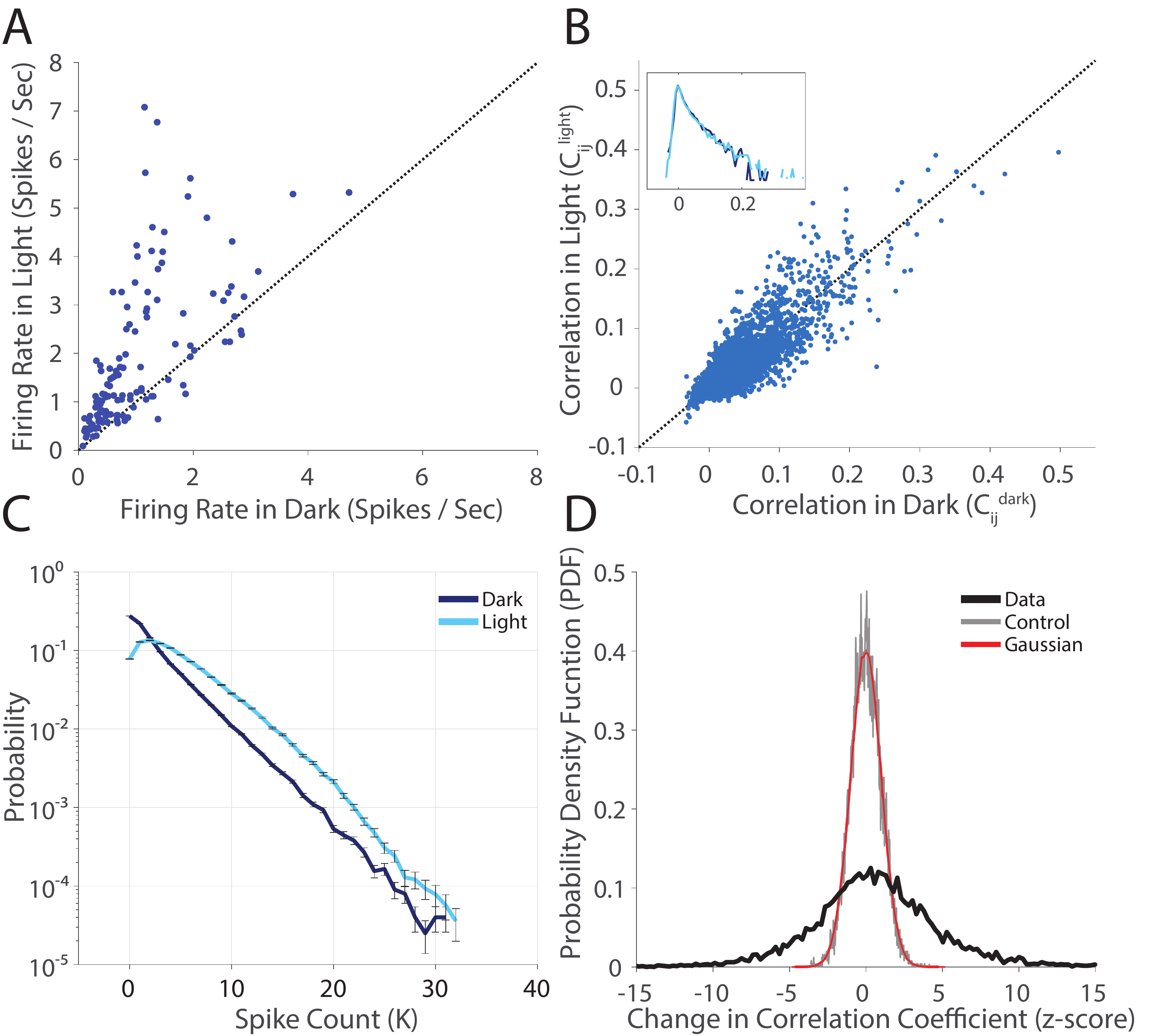}}
\caption{ {\bf Statistically Significant Changes in Population Structure Across Stimulus Conditions}. {\bf A}. Firing rate in the {\it light} condition plotted against firing rate in the {\it dark} for all $N =128$ cells. Error bars are given by the standard error of the mean (SE) and are smaller than the plotted points. {\bf B}. Pairwise correlation in the {\it light} plotted against pairwise correlation in the {\it dark}, for all $N=128 \cdot 127 / 2$ pairs of cells. Error Bars are not shown (but see panel {\bf D}). Inset is the probability density function (PDF), on a log scale, of correlation coefficients. {\bf C}. Measured P(K) in the two light conditions. $K$ is the number of active cells in a state. Error bars are given by the SE. {\bf D}. The PDF of z-scores of changes in correlation coefficients. The change in correlation coefficient is normalized by the error ($\sigma =  \sqrt{ \sigma_{d}^2 + \sigma_{l}^2 }$). These error bars are standard deviations over bootstrap resamples of the data, estimated per cell pair. Data compares the {\it light} and {\it dark} adapted conditions (thick black line), the control compares a random half of the dark dataset to the other half (gray), and a numerical gaussian is plotted in red for comparison.}
\label{fig:fig2}
\end{figure}

We found that firing rates mostly increased at the higher light level, an effect that was highly statistically significant (Fig.  \ref{fig:fig2}A). The correlation coefficients between pairs of cells showed some increases and some decreases between the {\it light} and {\it dark} adaptational states (Fig. \ref{fig:fig2}B). Overall, the distribution of correlation coefficients was roughly the same. However, the detailed pattern of correlation appeared to change. To evaluate the significance of changes in correlation coefficients, we estimated the difference between the correlation coefficients in the two light-adapted conditions normalized by the error bar - a measure also known as the z-score. The distribution of z-scores across all pairs of ganglion cells (Fig. \ref{fig:fig2}D, black curve) had significant density in the range of $5-10$ standard deviations, a result that is not consistent with controls (random halves of the {\it dark} dataset compared again each other, gray curve), or with the curve expected for the null hypothesis (a Gaussian with standard deviation of one, red curve). Thus correlation coefficients between individual pairs of cells change far more than expected by chance. The final ingredient in the maximum entropy model, the probability of {\it K} simultaneously active cells, ({\it P(K})), displayed a statistically significant overall shift towards sparseness in the {\it dark} adapted condition (Fig. \ref{fig:fig2}C).

Following previously published results \cite{tkacik2014, tkacik2015}, we modeled the distributions of neural activity with the k-pairwise maximum entropy model, which approximates the probability of all patterns of activity in the ganglion cell population. Here, we binned each ganglion cell's spike train in 20 ms time windows, assigning 0 for no spikes and 1 for one or more spikes, $r_i = [0,1]$. We denote a particular population activity pattern as {\it R} $=\{ r_i \}$. The probability of state {\it R} in this model is given by:

\begin{equation}
P(R) = \frac{1}{Z} \exp( -E_{\mathrm{k-pairwise}}(R))
\end{equation}

where {\it Z} is the normalizing factor, and we've introduced a unitless `energy': 

\begin{equation}
E_{\mathrm{k-pairwise}}(R) =  \sum_i^N h_i r_i + \sum_{i,j \neq i}^N J_{ij} r_i r_j + \sum_{k=1}^K \delta \left (  \sum_i r_i,  k \right ) \lambda_k
\end{equation}

\noindent with $\delta$ the Kronecker delta. The shapes of the probability and energy landscapes have a one-to-one relationship to each other, with energy minima corresponding to probability maxima. Because of the extensive intuition surrounding the concept of an energy landscape in physics, we will often use this term. This model is constrained to match the expectation values $\langle r_{i}\rangle$, $\langle r_{i}r_{j}\rangle$ and {\it P(K)} measured in the data. We inferred these models with a modified form of sequential coordinate descent \cite{dudik, broderick, tkacik2014} [see also Methods, Fig. S2-5].

\subsection*{The Collective State of a Neural Population}

A first step towards the study of collective phenomena in neural populations is to understand what is the qualitative nature or `phase' of the neural population. Phases of matter occur everywhere in nature where there is some collective structure in the population. First-order phase transitions can occur when a particular system can decrease its free energy by transitioning to a new phase. Thus, an explanation of previously observed phase transitions is that the pattern of correlation among ganglion cells induces a highly structured phase which is qualitatively different from the phase found in the high temperature limit. From this perspective then, we ask whether this phase is robust to different adaptational states, and what are the properties of the retinal population code that give rise to that phase? 

\begin{figure}[t]
\centerline{\includegraphics[width=0.44\textwidth]{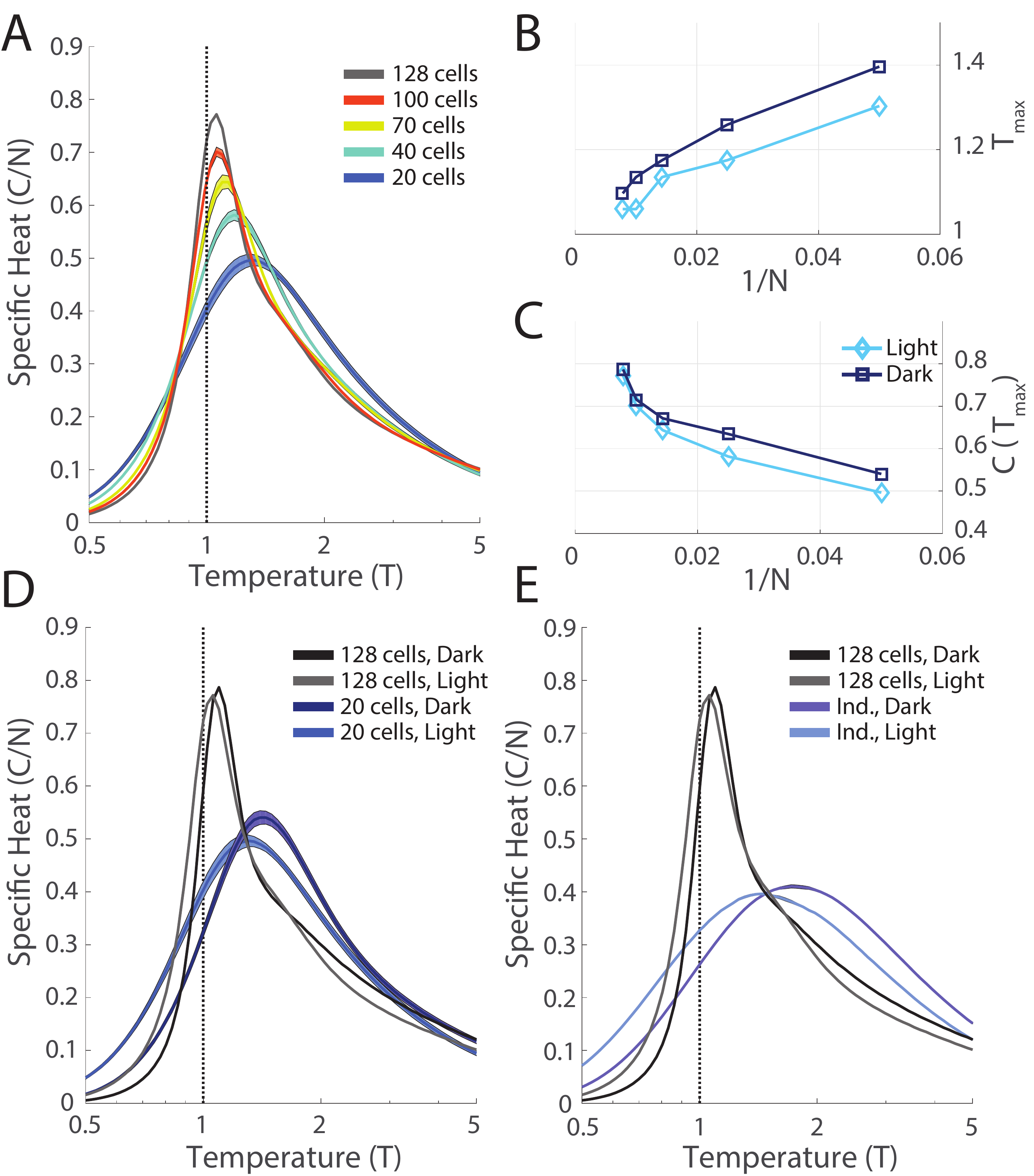}}
\caption{{\bf Phase Transitions Robustly Present in Both Stimulus Conditions}. {\bf A}. For each value of {\it N}, we selected 10 groups of cells (out of the {\it N}=128, M1, {\it light} recording). For each group we inferred the k-pairwise maximum entropy model as in \cite{tkacik2015}, and estimated the specific heat. Shaded areas are the SE. {\bf B,C}. Peak temperature, $T_{max}$, and peak specific heat, $C(T_{max})$, plotted as a function of inverse system size, $1/N$. {\bf D}. Specific heat plotted versus temperature for the full network ($N=128$) and for smaller subnetworks ($N=20$) for both the {\it dark} and {\it light} conditions. Error bars as in {\bf A}. {\bf E}. Comparison of the specific heat of the full network to that of an independent network estimated from shuffled data, for both the {\it dark} and {\it light} conditions.}
\label{fig:fig3}
\end{figure}

To study the emergence of a phase transition with increasing system size, we subsampled groups of {\it N} neurons and inferred models for these subsets of the full neural data. For all of these networks, we then introduced a fictitious temperature parameter, {\it  T}, into the distribution, $P(R) = (1/Z(T)) \exp(-E(R)/T)$. This parameter allows us to visit parameter regimes of our model where the qualitative nature of the system changes. If the shape of the specific heat as a function of temperature exhibits a sharp peak, this indicates a phase transition - a macroscopic restructuring of the properties of the system across parameter regimes. Thus, an analysis where we vary the effective temperature allows us to gain insight into the state of the real neural population at {\it T}=1. As previously described \cite{tkacik2015}, we found a peak in the specific heat that sharpened and moved closer to {\it T}=1 with increasing system size, {\it  N} (Fig. \ref{fig:fig3}A-C).

The systematic changes that we observe as a function of the system size {\it N} (Fig. \ref{fig:fig3}A-C) indicate that correlation plays a more dominant role as the population size increases. To further understand the role of correlations, we performed a shuffle test, where we broke correlations of all orders in the data by shifting each cell's spike train by a random time shift (including periodic wrap-around) that was different for each cell. Following this grand shuffle, we repeated the full analysis procedure described above (fitting a maximum entropy model to the shuffled data and estimating the specific heat). We found that the heat capacity had a much lower and broader peak that did not change as a function of {\it N}. In addition, this heat capacity curve agreed closely with the analytical form of the specific heat for an independent neural population (Fig. S6). This analysis demonstrates that the sharpening of the specific heat that we observed is a direct consequence of the measured pattern of correlation among neurons [Fig. \ref{fig:fig3}E].

The shuffled curves were noticeably different across light adapted conditions (Fig. \ref{fig:fig3}D). This is not surprising as the analytical form for the specific heat of a network of independent neurons depends only on the average firing rate of each neuron, and these are substantially different between the two luminance conditions (Fig. \ref{fig:fig2}A). However, the heat capacity peaks for both the {\it dark} and the {\it light} conditions became more similar with increasing {\it N}. Clearly some macroscopic properties of the network were conserved across luminance conditions for the real, correlated, data (Fig. \ref{fig:fig3}).

\begin{figure}
\centerline{\includegraphics[width=0.5\textwidth]{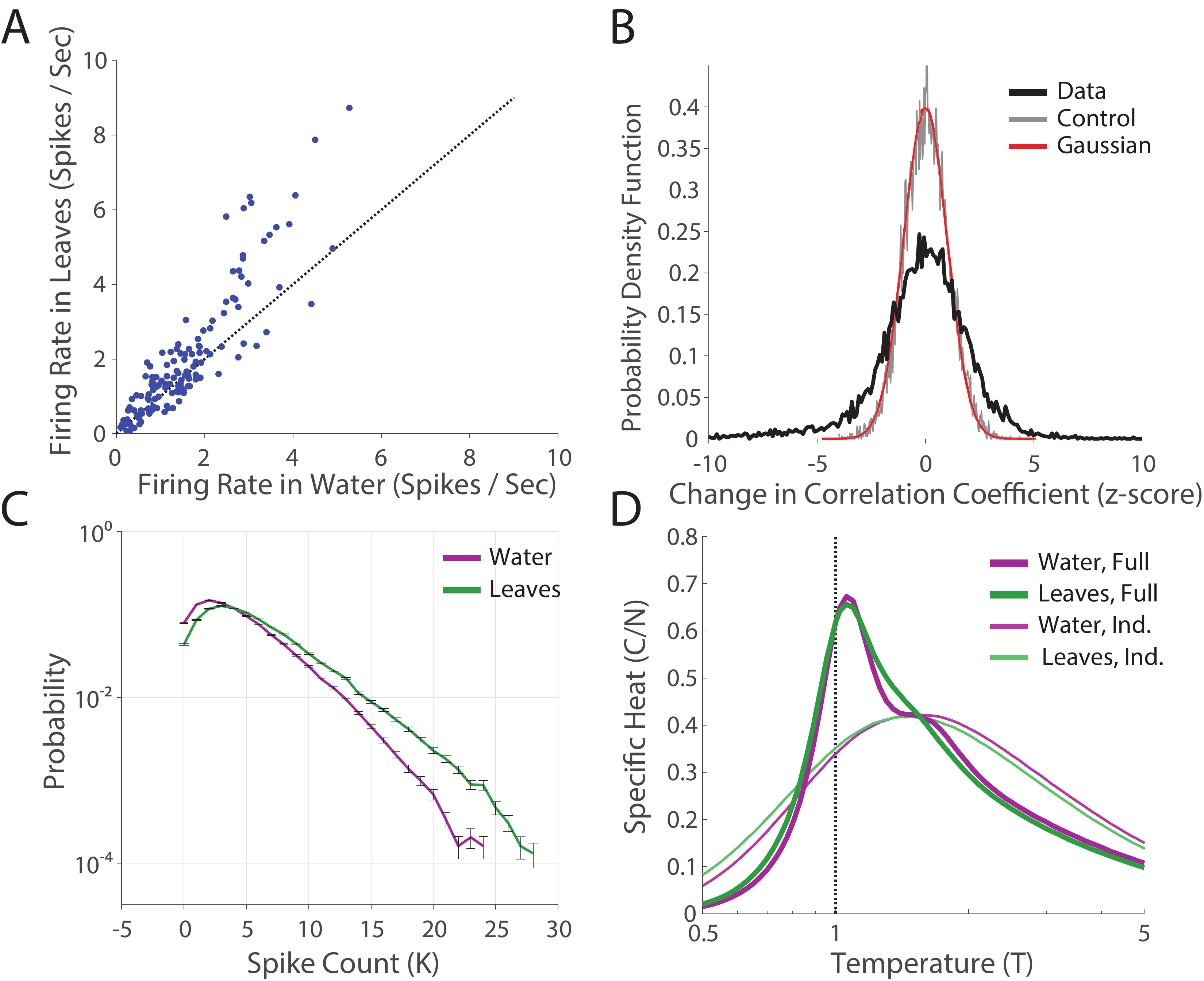}}
\caption{ {\bf Robustness of Phase Transition to Alternate Adaptive Mechanisms}. {\bf A}. Firing rate  in the ``leaves'' natural movie (M1) plotted against firing rate in the ``water'' natural movie (M2) condition, for all $N=140$ cells. Error bars are the SE. {\bf B}. The distribution of changes in correlation coefficient across the two stimulus conditions, for all $N=140 \cdot 139 / 2$ pairs of cells. {\bf C}. Distributions of the spike count, P({\it K}), across the two stimulus conditions. {\bf D}. Specific heats for full data and shuffled data (calculated as in Fig. 3), in the two natural movie conditions. }
\label{fig:fig4}
\end{figure}

The correlation structure of natural movies can in principle trigger a broad set of observed retinal adaptation mechanisms, such as adaptation to spatial contrast \cite{smirnakis, baccus2002}, temporal contrast \cite{chander}, and relative motion of objects \cite{olveczky}. To generalize our results to these higher-order adaptive mechanisms, we ran another experiment comparing the distributions of responses of the same retina to two different natural stimuli ensembles, without a neutral density filter (Experiment $\#2$, see Methods). These two natural movies were of grass stalks swaying in the wind (M1, the same movie as in the previous experiment), and ripples on the surface of water near a dam (M2). The first movie (M1) had faster movements, larger contrasts, and fewer periods of inactivity. Likely as a consequence, we found higher firing rates in ganglion cells during M1 (Fig. \ref{fig:fig4}A). We found statistically significant differences in the correlation coeffecients, $C_{ij}$,  and {\it P(K)} across the two stimulus conditions (Fig. \ref{fig:fig4}B,C). However, the specific heats of the full networks in the two movies sharpened similarly across conditions (Fig. \ref{fig:fig4}D), indicating that this macroscopic property of the retinal population code was also robust to different choices of naturalistic stimuli. 

\subsection*{The Role of Retinal Processing for the Collective State}

So far, our results have demonstrated that the peak in the specific heat is due to the pattern of correlation among neurons. However, these correlations have contributions both from retinal processing, such as the high spatial overlap between ganglion cells of different functional type \cite{segev2004, puchalla}, and from the correlation structure in the stimulus itself. In order to compare the relative importance of these two different sources of correlation among ganglion cells, we measured neural activity during stimulation with a randomly flickering checkerboard. By construction, our checkerboard stimulus had minimal spatial and temporal correlation: outside of 66 $\mu m$ squares and 33 ms frames, all light intensities were randomly chosen. Returning to Experiment $\#1$ in the light-adapted condition, we compared the response of the retina to the natural movie and the checkerboard stimuli (Note that here we are working with the $N=111$ ganglion cells that were identifiable across both conditions, a subset of the $N=128$ ganglion cells we worked with in Figure \ref{fig:fig3}). 

\begin{figure}
\centerline{\includegraphics[width=0.5\textwidth]{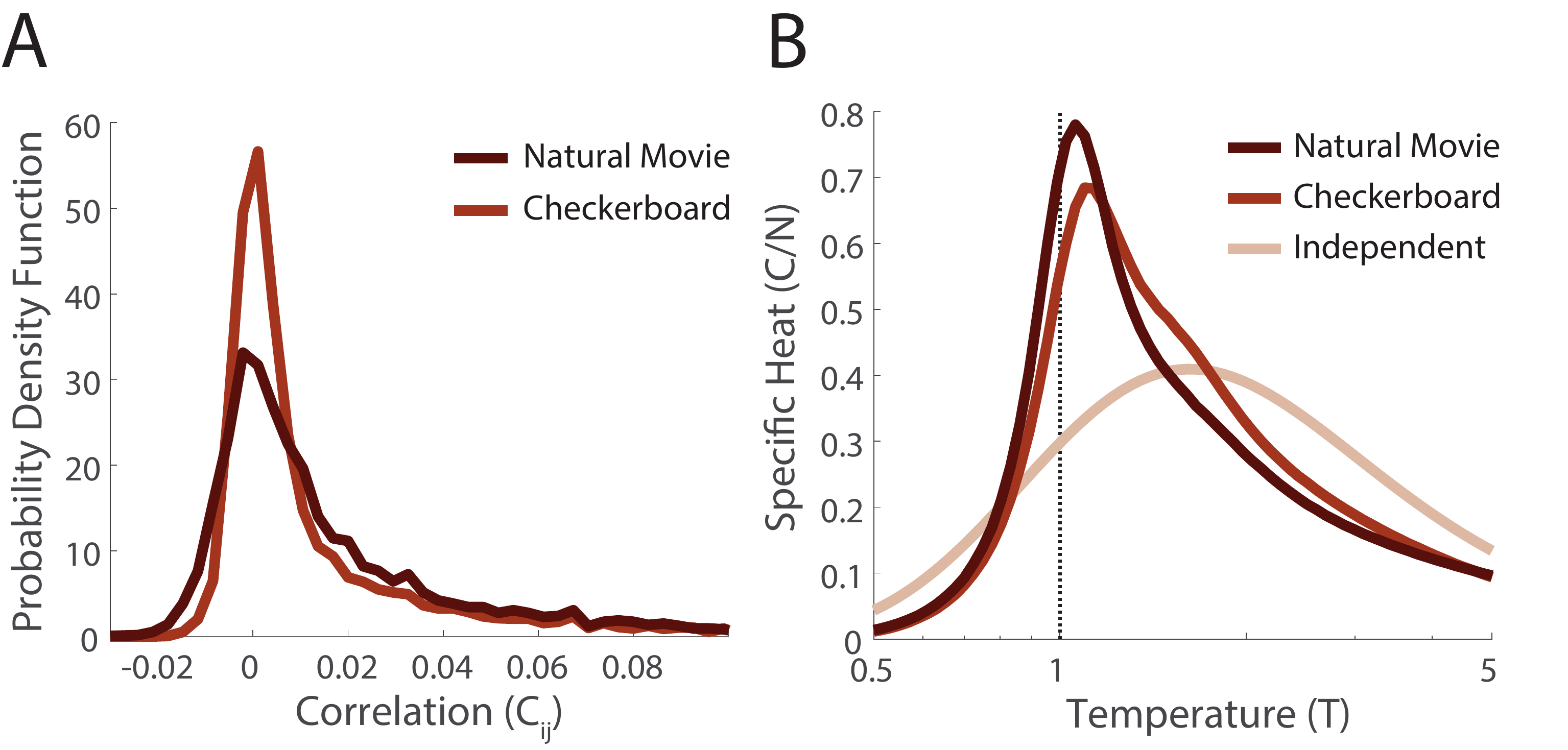}}
\caption{{\bf Dependence on Correlation Structure of Naturalistic Stimuli}. {\bf A}. Distributions of correlation coefficients in the naturalistic and artificial stimulus conditions (Experiment $\# 1$, {\it light}, $N=111$ ganglion cells). {\bf B}. Specific heats in the naturalistic and artificial stimulus conditions. Here the independent curve was calculated analytically based on the firing rates in the checkerboard condition.}
\label{fig:fig5}
\end{figure}

The distribution of pairwise correlation coefficients was tighter around zero when the population of ganglion cells was responding to the checkerboard stimulus (Fig. \ref{fig:fig5}A). The specific heat in the checkerboard was smeared out relative to the natural movie, but was still very distinct from the independent population (Fig. \ref{fig:fig5}B). This suggested to us that most, but not all, of the contributions to the shape of the specific heat were shared across the two stimulation conditions, and therefore arose from retinal processing.

A simple and popular view of retinal processing is that each ganglion cell spike train is described by the spatio-temporal processing of the cell's classical receptive field. In this picture, correlation between ganglion cells arises largely from common input to a given pair of ganglion cells which can be described by the overlap of their receptive fields. To explore the properties of this simple model, we estimated linear-nonlinear (LN) models for each of the ${\it N}=111$ ganglion cells in the checkerboard recording (Methods). We then generated spike trains from these model neurons responding to a new pseudorandom checkerboard sequence, and binarized them into 20ms bins in the same manner as for the measured neural data. As expected, the receptive fields had a large degree of spatial overlap \cite{segev2004, segev2006}, which gives rise to significant stimulus-dependent correlations.  

\begin{figure}
\centerline{\includegraphics[width=0.5\textwidth]{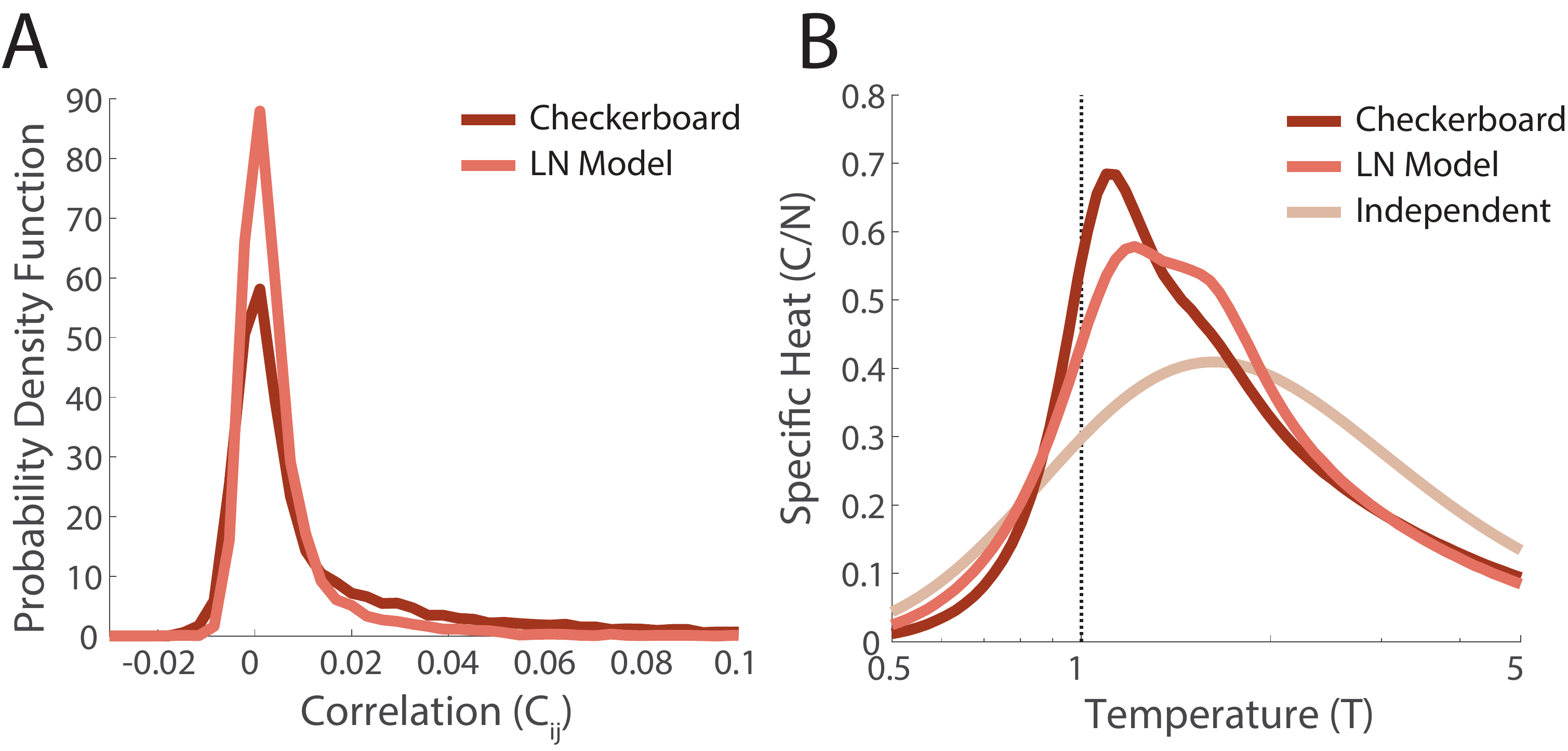}}
\caption{{\bf Networks of Model LN Neurons}. Model LN neurons were fit to the measured receptive fields as described in the text, $N=111$ ganglion cells. {\bf A}. Distributions of correlation coefficients over pairs of cells estimated in the training data (responding to checkerboard), and the simulated network of LN neurons. {\bf B}. The specific heats in the checkerboard and simulated LN network. The independent curve is the same analytic estimate as in Figure 5.}
\label{fig:fig6}
\end{figure}

We found that these networks did not reproduce the distributions of correlations found in the data, instead having lower values of correlation and fewer outliers (Fig. \ref{fig:fig6}A). The specific heat of the network of LN neurons was reduced relative to the neural data that the LN models were based upon (Fig. \ref{fig:fig6}B). Thus, the peak in the specific heat is enhanced by the nonlinear spatial and temporal computations in the retina that are not captured by models of the classical receptive field.

\subsection*{What Pattern of Correlation is Needed for the Collective State?}

Because the detailed properties of the maximum entropy model depend strictly on correlations measured within the neural population, we wanted to develop a more general understanding of what aspects of the pattern of correlation were essential. To do this, we altered particular properties of the measured matrix of correlations, keeping the firing rates constant. We then inferred the maximum entropy model parameters for these new sets of constraints, and estimated the specific heat. For these manipulations, we worked with the simpler pairwise maximum entropy model. We made this choice for several reasons. First, manipulating only the pairwise correlation matrix made our analysis simpler and more elegant than also having to perturb the distribution of spike counts, {\it P(K)}. There is a large literature reporting values of pairwise correlation coefficients, helping us to make intuitive choices of how to manipulate the correlation matrix, while very little such literature exists for {\it P(K)}. Additionally, any perturbation of the correlation matrix consequently changes {\it P(K)}, so that attempting to change the correlation matrix while keeping {\it P(K)} fixed is a nontrivial manipulation. Second, in the pairwise model all effects of correlational structure are confined to the interaction matrix. This interaction matrix has been studied extensively in physics \cite{fischerhertz}, and hence there is some intuition as to how to interpret systematic changes in the parameters. Conversely, we have little intuition currently for the nature of the k-potential. Our final and most important reason was that the qualitative behavior in the heat capacity (sharpening with system size, convergence across light and dark datasets) is the same for both pairwise and K-pairwise models across all conditions tested (Fig. S7).

\begin{figure}
\centerline{\includegraphics[width=0.5\textwidth]{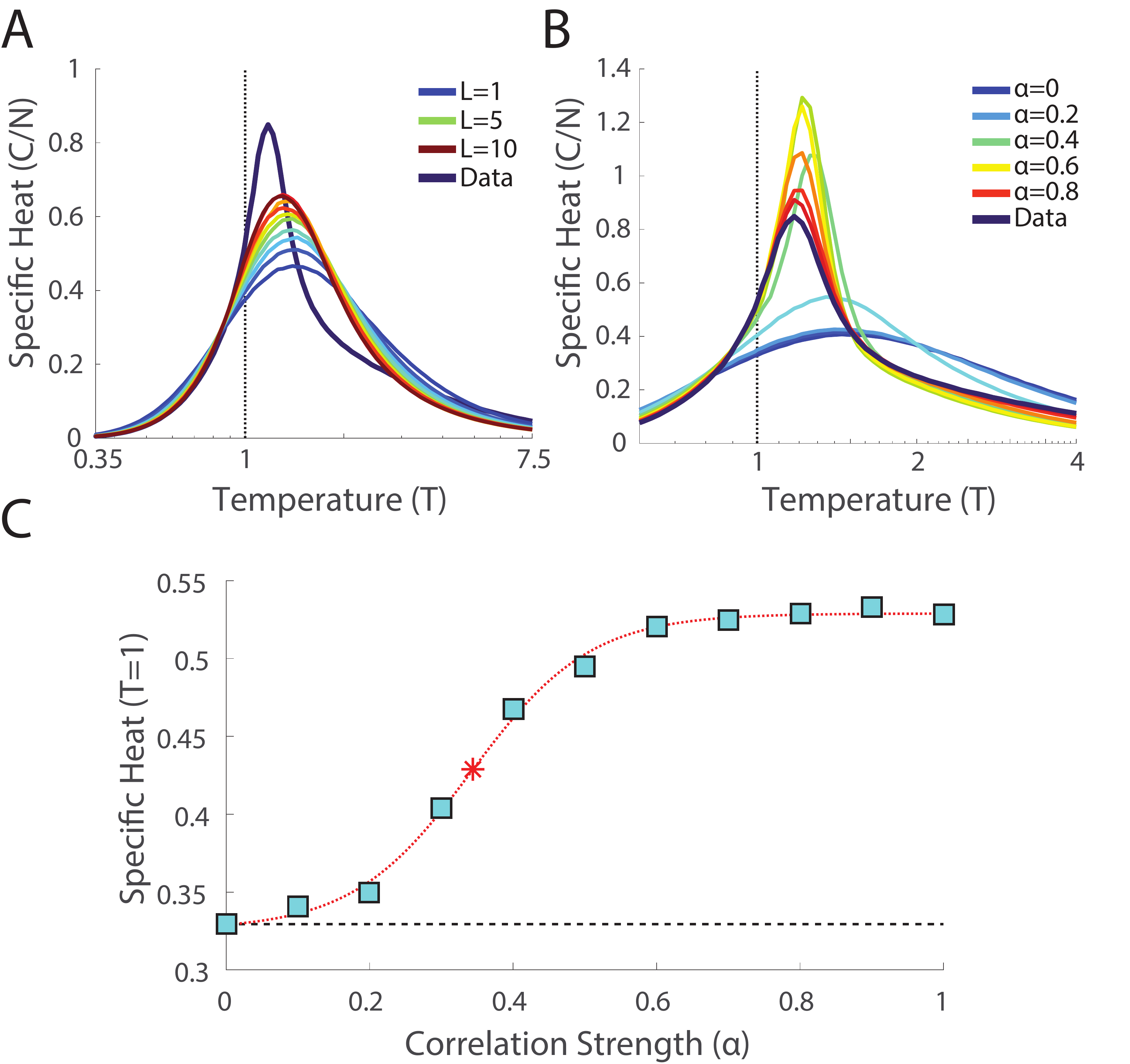}}
\caption{{\bf Targeted Manipulations of the Correlation Matrix}. {\bf A}. The specific heat plotted versus temperature for several values of the connectivity, $L$ (see text). {\bf B}. The specific heat plotted versus temperature for several values of the global covariance $\alpha$ (see text). {\bf C}. The specific heat at the operating point of the network ({\it T}=1) plotted as a function of $\alpha$. In red, a logistic fit (MATLAB's nlinfit of the form $y(x) = A + B \cdot (\exp{C (x-D)} + 1)^{-1}$). The asterisk denotes the halfway value ($\alpha^* = 0.34$). Black dashed line gives the value for an independent network. }
\label{fig:fig7}
\end{figure}

The correlation matrix in the retinal population responding to a natural stimulus has many weak but statistically non-zero correlations \cite{puchalla, segev2006}, a result also found elsewhere in the brain \cite{cohenkohn, reichvictor}. To test their contribution to the specific heat, we kept only the largest {\it L} correlations per cell, replacing the other terms in the correlation matrix with estimates from the shuffled (independent) covariance matrix. If our results are based on a ``small world network'' of a few, strong connections \cite{barabasi}, then the specific heat for small values of {\it L} should begin to approximate our results for the real data. Clearly (Fig. \ref{fig:fig7}A), even keeping the top $L=10N$ (out of a total of $L=63.5N$ values) strongest correlations did not reproduce the observed behavior. Therefore, the full ``web'' of weak correlations contributed substantially to the shape of the specific heat of the retinal population code.

We were next interested in understanding the qualitative nature of how networks transition between the independent and fully correlated regimes. Our approach was to scale all the correlations down by a constant ($\alpha$), and to follow the inference procedure described above. Specifically, we formed a new correlation matrix $C_{ij}^{\,\mathrm{mixed}}(\alpha)=\alpha C_{ij}^{\, \mathrm{true}}+(1-\alpha)C_{ij}^{\, \mathrm{shuff}}$. We found that the specific heat of the neural population exhibited a sharp transition between independent and fully correlated behavior, as the correlation strength, $\alpha$, ranged from $0$ to $1$ (Fig. \ref{fig:fig7}B). Phase transitions similar to the one observed in the full model emerge roughly at a critical value of the correlation strength, $\alpha^* =  0.34$ (Fig. \ref{fig:fig7}C). Interestingly, $\alpha^*$ was substantially smaller than the measured correlation strength, $\alpha=1$.  This indicates that the population of retinal ganglion cells had a overall strength that was ``safely'' within the strongly correlated regime.  Thus, the low temperature state is robust to changes in adaptational state or stimulus statistics that might shift the overall strength of correlations among neurons.

 \subsection*{What is the Nature of the Collective State?}

Our hypothesis was that the emergence of a phase transition was correlated with the emergence of structure in the energy landscape. Previously, the structure of the energy landscape has been studied with zero temperature Monte Carlo (MC) mapping of local minima \cite{tkacik2014}, where one changes the activity state of single neurons such that the energy of the population activity state always decreases. States from the data were thereby assigned to local minima in the energy landscape, which can be thought of as a method of clustering a set of neural activity patterns into population ``codewords'' \cite{tkacik2014}. If each cluster encodes a specific visual stimulus or class of stimuli, then this clustering operation provides a method of correcting for errors introduced by noise in the neural response.

There are two reasons why we chose to study the structure of the energy landscape at the operating point of the system ({\it T} =1). First, when we performed zero temperature descent with our models, our primary finding was that the overwhelming majority of states descended into the silent state (only $503$ out of $1.75 \cdot 10^5$ did not descend into silence on a sample run). This indicated that the energy landscape had very few local minima. Thus we needed a different approach to explore the structure of the energy landscape. Second, we were interested in properties of the system (such as the specific heat) that were themselves temperature dependent, so it made sense to stick with the real operating point of the neural population ({\it T} =1).

When analyzing sufficiently large neural populations (typically, ``large'' means $N > 20$ cells), there are too many states to simply ennumerate all of them.  As a consequence, the energy landscape was accessed indirectly, through a Markov Chain Monte Carlo sampler \cite{mackay}, which simulates an exploration of phase space by defining the state-dependent transition probabilities between successive states. Provided that these transition probabilities are properly defined, the distribution of samples drawn should approach the desired (true) distribution with sufficient sampling. The set of these transition probabilities across all the neurons defines a `direction of motion' in neural response space. We will study these directions of motion as a way to gain more insight into the properties of the energy landscape of our measured neural populations.

To study the relationship between the directions of motion given by the MC sampling process and the observed phase transition, we returned to the manipulation with scaled covariances. For a given state {\it R}, the Monte Carlo sampler in each model (inferred for a particular value of the correlation strength $\alpha$) will return a vector of conditional probabilities $X (R , \alpha) = \{ x_i \}$, where the conditional probability for each cell $i$'s activity is given by 

\begin{equation}
x_i (R , \alpha) = \mathrm{exp}(h_i^{(\mathrm{eff},\alpha)}  (R )) / ( \mathrm{exp}(h_i^{(\mathrm{eff},\alpha)} (R )) +1)
\end{equation}

with an effective field, given by

\begin{equation}
h_i^{(\mathrm{eff}, \alpha)}  (R ) = h_i^{(\alpha)} + \sum_{j \neq i} J_{ij}^{(\alpha)} r_j 
\label{eq:efffield}
\end{equation}

\begin{figure}
\centerline{\includegraphics[width=0.5\textwidth]{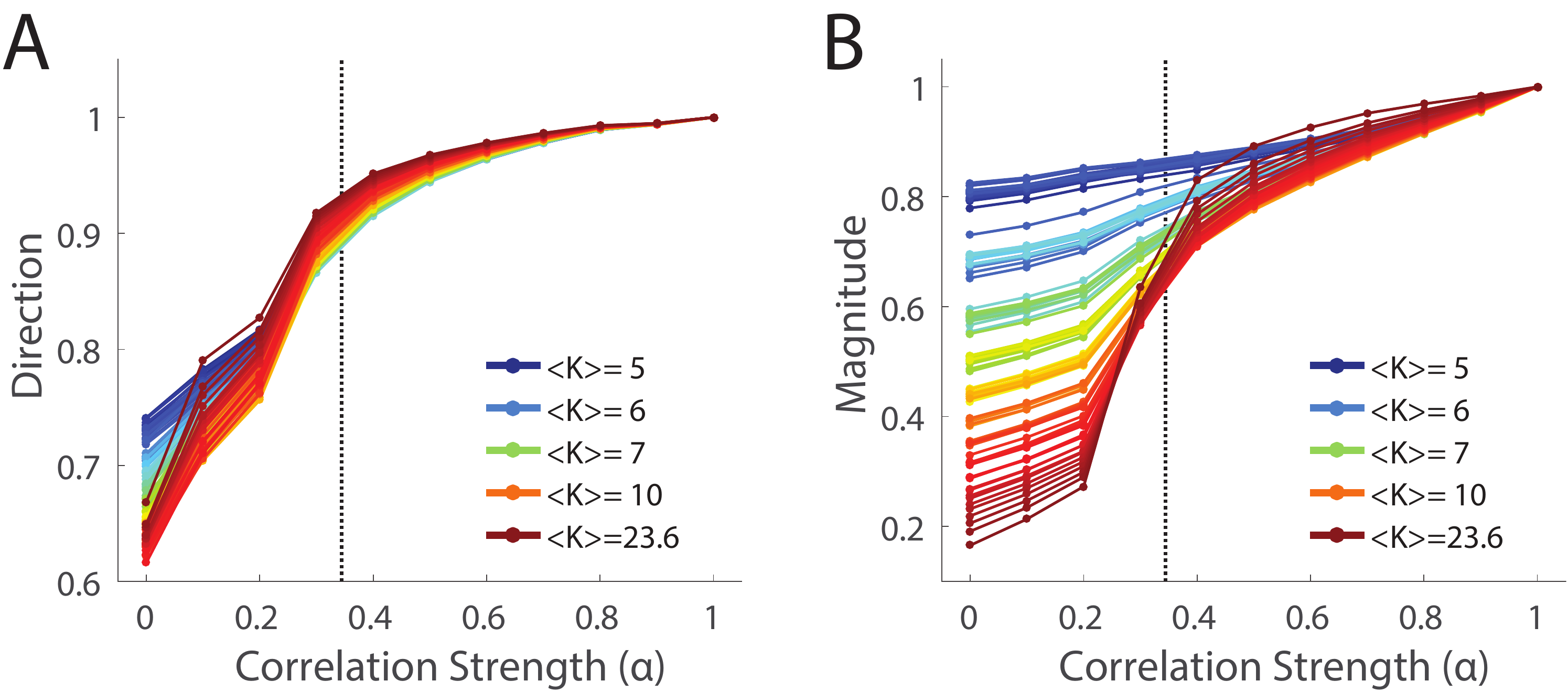}}
\caption{{\bf Emergence of Structure in the Energy Landscape at Low Covariance Strength}. All states in the {\it light} dataset with five or more spike counts were sorted by spike count and grouped into 100 equally populated groups. Averages were calculated over these groups of states (for clarity indexed by $p$); the color of the line indicates the average spike count within a group ($1613$ states per group). {\bf A}. Average similarity in direction of sampling between full model and model at correlation strength $\alpha$. This is estimated as the average over $p$ data states: $ \langle \hat{X}( R_p, \alpha) \cdot \hat{X}( R_p, \alpha {=}1) \rangle_p$.  {\bf B}. Similarities of magnitudes, estimated as $\langle \, \mathrm{RMS} (  X( R_p, \alpha) )  \, /  \, \mathrm{RMS} ( X( R_p, \alpha{=}1) ) \, \rangle_p$. The black dotted line in both panels is the halfway point $\alpha^*$, estimated from the logistic fit in Fig. 7 .}
\label{fig:fig8}
\end{figure}

We can now ask how the shape of the energy landscape evolves with respect to $\alpha$. Specifically, we compared the similarity in direction or magnitude of these `Monte Carlo flow' vectors with the vectors defined for the full model (at $\alpha=1$). We found large changes during the initial increase from $\alpha = 0$ to the critical correlation strength $\alpha^*=0.34$. At $\alpha^*$, the directions of Monte Carlo flow were on average $90 \%$ similar to the fully correlated neural population (Fig. \ref{fig:fig8}A). Similarities in magnitude of these vectors also increased substantially by this point (ranging from $60 \%$ to $85 \%$, Fig. \ref{fig:fig8}B). Thus, the structure of the energy landscape was largely established by the transition point $\alpha^*$. From $\alpha^*$  onwards, changes in direction grew progressively smaller with increasing $\alpha$, while vector magnitude evolved smoothly all the way up to $\alpha=1$. This suggested that the `shape' of the energy landscape emerged at a correlation strength near $\alpha^*$ and that further increases in $\alpha$ served to `deepen' the existing contours in the energy landscape.

\begin{figure}[h]
\centerline{\includegraphics[width=0.5\textwidth]{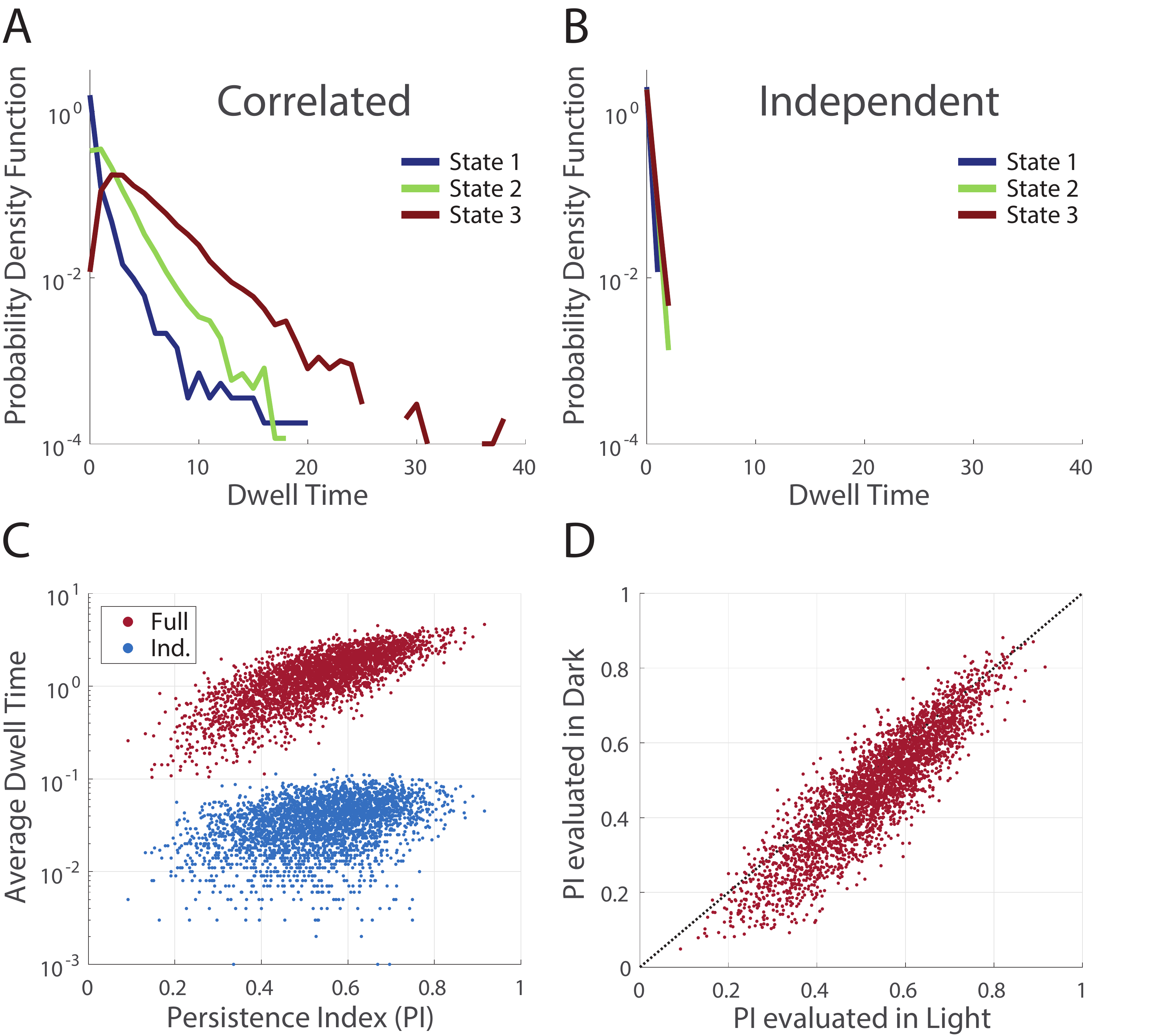}}
\caption{{\bf Visualizing the Structure of the Models Through the Dwell Times in Sampling}. {\bf A}. Distributions of dwell times for three sample states, estimated over $10^4$ separate instantiations of MC sampling from the full model. The persistence indices for states 1,2 and 3 were $0.092$, $0.54$, and $0.92$, respectively. {\bf B}. For the same states as in ({\bf A}), distributions of dwell times estimated on shuffled (independent) data. {\bf C}. Across the $N=3187$ states with $K=12$ spiking cells recorded in the data (M1, {\it light}) we measured the average dwell time (over $10^3$ MC runs) in the full (red) and independent (blue) models. These are plotted vs. the PI given by the full model. Note the logarithmic scale on  the y-axis. {\bf D}. The persistence indices for the same group of states are estimated using the maximum entropy model fitting the natural movie in the {\it light} (x-axis) and the {\it dark} (y-axis) adapted conditions. }
\label{fig:fig9}
\end{figure}

While our energy landscape does not have many true local minima, we can gain insight into the nature of the energy landscape induced by correlations by considering how long the system remains in the vicinity of a given state under $T=1$ MC dynamics. Since the experimentally measured neural activity is sparse, the directions of motion are heavily biased towards silence. Regardless of initial state, the sampler will eventually revisit the silent state. 

To demonstrate this, we returned to the data from Experiment $\#1$ (M1, {\it light}), and the corresponding k-pairwise model fits. Due to the additional constraints the effective field was now $h_i^{\mathrm{eff}}(R) = h_i + \sum_{j \neq i} J_{ij} r_j + \lambda_{K+1} - \lambda_K$,  with $K = \sum_{j \neq i} r_j$. We worked with all the states observed in the {\it light} condition that had $K=12$ spiking cells (total number of states = 3187 ). We selected a set of initial activity states all having the same value of {\it K}, as these analyses depended strongly on {\it K}. We chose a value of {\it K} large enough that effective fields were large, and hence collective effects of the population code were significant.  At the same time, we wanted {\it K} to be small enough that we could observe many such states in our sampled experimental data.

We measured the distributions of dwell times in the vicinity of a particular state by running many separate Monte Carlo runs initialized from the same initial state. On each of these runs, we defined the dwell time as the number of MC samples (where all N cells were updated) required to change 9 of the originally spiking cells to silent. For each given initial state, our analysis produced a distribution of dwell times, due to the randomized order of cell choice during sampling, as well as the stochasticity inherent in sampling at $T=1$. 

We found significant differences in the distributions of dwell times across different initial states (Fig. \ref{fig:fig9}A). This demonstrated that Monte Carlo flow was trapped for a longer amount of time in the vicinity of particular states. For the same initial states, average dwell times measured on the energy landscape for independent models were almost an order of magnitude shorter (Fig. \ref{fig:fig9}B). 

We searched for a measure that could capture the variability in dwell times across initial states in the full model, based on our intuition that a state that started near a local minimum would have a long persistence time before finite temperature MC sampling would move it far away. This led us to define a persistence index (PI) that captured the tendency of a state to remain nearby under MC sampling dynamics. Specifically, for a given state $R$, we define $PI = \hat {X}(R) \cdot \hat {R}$, namely the cosine of the angle between the initial state and the average next state. If the PI is close to 1, then the direction that the state {\it R} evolves towards under MC sampling is the state itself, and hence the state will remain the same. 

Over all the initial states studied, we found a large positive correlation between the average dwell times and the persistence indices (correlation coefficient of $0.75$, Fig \ref{fig:fig9}C). This significant correlation justifies the use of the PI as a simpler proxy for the dwell time. In contrast, the correlation was only 0.38 when measured on the energy landscape of the independent model and dwell times were systematically smaller by orders of magnitude (Fig. \ref{fig:fig9}C blue).

The persistence index also allowed us to characterize the similarity of the neural code for the same natural movie between light and dark adapted conditions. If the energy landscape changed between the {\it dark} and {\it light} conditions, then one would expect the dwell times to change as well. Instead, we found a strong correspondence across the {\it light} and {\it dark}  experimental conditions, that was absent in the independent model (Fig. \ref{fig:fig9}D, Fig. S8 correlation coefficient of $0.90$ vs $0.37$). To estimate the variability in this measure, we compared the PI across models inferred for two separate random halves of the {\it light} condition (Fig. S8, correlation coefficient of $0.97$). Thus, the pattern of correlation measured in the two luminance conditions created similar structure in the system's energy landscape, even though the detailed statistics of neural activity were different. This structure endows the population code with a form of invariance to light level that is not present at the level of individual ganglion cells.

\section*{Discussion}

Across different naturalistic stimulus conditions, we observed a similar sharpening of the specific heat as we increased the number of retinal ganglion cells, {\it N}, that were analyzed together. The size and shape of the peak in the specific heat depended on the pattern of correlation measured experimentally. The primary contribution to these macroscopic properties arose in retinal processing, with smaller contributions from correlations in the stimulus itself. A simple model of retinal processing (the LN model) captured some of these properties, but did not reproduce the results of our analysis of real neural populations. In the measured matrix of pairwise correlations between ganglion cells, the many weak but statistically non-zero correlations contributed substantially to the sharpening of the peak in the heat capacity. The overall correlation matrix can be scaled down by a factor of two without substantial changes to the structure of the energy landscape. The sharpening of the specific heat corresponds clearly with the emergence of structure in the energy landscape, as captured by an analysis of Monte Carlo sampling dynamics. This structure consists of regions which weakly ``trap'' the system during exploration of the energy landscape, akin to short lived attractors in a dynamical system.

The peak in the specific heat on its own is not sufficient to characterize the neural population, as qualitatively different probability distributions can have similarly sharp peaks in the specific heat. For this reason we've worked with distributions that are based on measured neural data, where the changes between distributions are known (as in our manipulation scaling the correlation globally). In such manipulations, systematic shifts in the sharpness of the peak in the specific heat correspond to systematic shifts in the structure of the low temperature phase. Thus while a peak in the specific heat does not provide a full characterization of the neural distribution, it can be a useful tool for comparing structure in similar networks.

In this study, we have chosen not to attempt to extrapolate to the infinite system size limit. Our dense multi-electrode arrays were designed to record from most of the retinal ganglion cells within a correlated patch, where there is a high probability of two neurons sharing significant correlation. For neurons more widely separated, the strength of correlation decreases \cite{puchalla}. Therefore, the trend that we observed in the specific heat versus the number of neurons analyzed within a correlated patch is expected to be different than the trend for ganglion cells with larger spatial separation, making a naive extrapolation unlikely to be accurate \cite{macke2016}.

A recent study tested the generality of phase transitions in simulations of retinal ganglion cells  \cite{macke2016}, simulating networks of neurons in the retina, and then estimating the specific heat following procedures similar to those presented here. Their results are consistent with ours: phase transitions are robustly present in different networks of neurons, and their presence is largely invariant to experimental details. Similar to us, they also found that the sharpness of the peak in the specific heat is systematically enhanced by ``stronger'' network correlations. Because Nonnenmacher et al. also found this behavior in very simple models, such as homogeneous neural populations, they concluded that the sharpening peak in the heat capacity does not provide a complete characterization of the population code.  We agree.  However, because we went on to analyze the structure of the probability landscape for our real, measured neural populations, we could show that the emergence of the peak was related to the clustering of neural activity patterns (Figs. 8 and 9). Another difference is that Nonnenmacher et al. focused on the large N limit of model populations, where any nonzero correlation will eventually result in a sharp peak.  Instead, we argue that it is more natural to focus on local populations of $N \sim 200$ ganglion cells, as these groups are likely to form the population coding units of the retina.

A separate line of work has investigated the presence of Zipf-like relationships in the probability distribution of neural codewords \cite{mora2011, schwab}.  A true Zipf Law is intimately related to a peak in the heat capacity at $T=1$, and Schwab et al. found that a Zipf Law was present under fairly generic circumstances, in which neural activity was determined by a latent variable (e.g., an external stimulus) with a wide range of values \cite{schwab}.  Again, this result is broadly consistent with our finding of great robustness of the low temperature state.  However, we have not chosen to analyze Zipf-like relationships in our experimental data for several reasons: 1) we can only sample $\sim 2$ orders of magnitude in rank (Fig. S9), making it difficult to estimate power law exponents; 2) we typically observe small deviations from the power law trend, and we are uncertain about how to interpret the importance of these ``bumps''. In addition, Schwab et al. focused on the large N limit, and their analysis is difficult to relate to the finite population sizes we considered here.

More generally, there are two types of phase transitions that are potentially consistent with the observed sharpening in the specific heat. The possibility that this phase transition is second-order, and hence associated with critical phenomena in the neural population code, was explored in the previous literature \cite{tkacik2015, mora2015, mora2011}. Here, we focused on the first-order phase transition in part because such phase transitions are far more common in nature, and hence offer the simplest possible explanation. Additionally, arguments for the critical tuning of neural population codes are strongly enhanced by evidence for optimization of the specific heat with respect to properties of the correlations. We did not find evidence for such optimality in our results. For example, when correlations were scaled down by a factor of 2 the specific heat remained essentially constant. Finally, this interpretation predicted that the emergence of a sharp peak in the specific heat would be strongly connected to the emergence of structure in the energy landscape, which our subsequent analyses demonstrated.

In our study, we have characterized the neural response in a single time bin, ignoring the role of temporal correlations across time bins. One can extend our approach to include temporal correlations by concatenating multiple time bins into each neural codeword. When the number of total time bins was systematically increased in such a manner, the peak in the specific heat sharpened substantially \cite{mora2015}. The authors interpreted these results as further evidence in favor of the critical properties of neural population codes. However, in all cases in both our study and of \cite{mora2015}, the peak was above $T=1$, consistent with our interpretation that neural populations are in a low temperature state. Since we take the sharpness of the specific heat to be an indicator of structure in the distribution of responses, this treatment of temporal correlations increases the structure of the low temperature state in a manner similar to an increase in the number of neurons analyzed together, {\it N}.

Using several different analysis methods, neural activity evoked by repeated presentations of the same stimulus has been shown to form clusters in the space of all possible activity patterns. Zero temperature descent in the energy landscape defined by the maximum entropy model mapped a large fraction of all neural activity patterns to non-silent energy basins, which were robustly activated by the same visual stimulus \cite{tkacik2006, tkacik2014}. Mapping neural activity patterns to latent variables inferred for a hidden Markov model revealed similar robust activation by the stimulus \cite{prentice, loback}. Huang et al. found a form of first-order phase transition as a function of the strength of an applied external field, from which they concluded that the energy landscape formed natural clusters of neural activity with no applied field \cite{huang}. Ganmor et al. recently uncovered a striking block diagonal organization in the matrix of semantic similarities between neural population codewords \cite{ganmor}, arguing for a clustered organization of neural codewords. All these analyses are likely to be different ways to view the same underlying phenomenon, although a detailed exploration of the correspondences among these methods is a subject for future work.

Are our results specific to the retina? In our approach, the collective state of the retinal population code is entirely determined by the pattern and strength of measured correlation. There is nothing about this pattern of correlation that makes specific reference to the retina. This means that any neural population having similar firing rates and pairwise correlations would also be in a similar collective state. Additionally, the strength of pairwise correlations we report here is smaller than or comparable to those reported in higher order brain areas, such as V1 and MT \cite{cohenkohn, reichvictor, berkes, okun}. This suggests that the collective state of neural activity, which arises due to a clustering of neural activity patterns, could occur throughout higher-order brain regions in population recordings of a suitable size ({\it N} $>$100).

\section*{Methods}

\subsection*{Experimental Recordings}

We recorded from larval tiger salamander ({\it Ambystoma tigrinum}) retina using the dense (30 $\mu$m spacing) 252-electrode array described in \cite{marre}. In Experiment $\#1$, which probed the adaptational state of the retina at normal and low ambient illumintation levels, the salamander was kept in dim light and the retina was dissected out with the pigment epithelium intact, to help the retina recover post dissection and adjust to the low ambient light levels in the {\it dark} condition. The rest of the procedure in Experiment $\#1$, and the full procedure for Experiment $\#2$, followed \cite{marre}.

\subsection*{Stimuli}

The chromatic checkerboard stimulus (CC) consisted of a random binary sequence per color (R,G,B) per checker, allowing 8 unique values for any given checker. Checkers were 66 $\mu$m in size, and refreshed at 30 Hz. There were two (gray scale, 8 bit depth) natural movies used: grass stalks swaying in a breeze (M1, 410 seconds) and ripples on the water surface near a dam (M2, 360 seconds). Both were gamma corrected for the CRT, and displayed at 400 by 400 pixel (5.5 $\mu$m  per pixel) resolution, at 60 Hz.

\subsection*{Experimental Details}

In Experiment $\#1$, after adapting the retina to the absolute dark for 20 minutes, we recorded in the {\it dark} condition first (by placing an absorptive neutral density filter of optical density 3 [Edmund Optics] in the light path), stimulating with (CC) for 60 minutes, and with (M1) for 90 minutes. The filter was then switched out for the {\it light} condition, in which we recorded for an additional 60 minutes of (M1) and another 60 minutes of (CC). To avoid transient light adaptation effects we removed the first 5 minutes of each recording (10 minutes from the first checkerboard) from our analysis. During stimulation with (M1) we sampled 340 sec long segments from (M1) with start times drawn from a uniform distribution in the [0 60] second interval of (M1). The spike sorting algorithm \cite{marre} was run independently on the recordings in response to (M1) at the two light levels, generating separate sets of cell templates at the two light levels, which were then matched across the two conditions, yielding $N=128$ ganglion cells. The spike trains were then binned in 20 ms time bins and binarized, giving $2.5 \cdot 10^{5}$ states in the {\it dark} and $1.75 \cdot 10^{5}$ states in the {\it light}. For the recordings from the checkerboard (in both light conditions), the templates from the {\it light} recording were used to fit the electrode activity. Across all four stimulus conditions this left us with $N=111$ cells for the comparisons of natural movies to checkerboard. The checkerboard in the {\it light} condition had $1.3 \cdot 10^5$ states.

In Experiment $\#2$, we alternated stimulation between (M1) and (M2) every 30 seconds, sampling 10 sec segments from both movies. For our analysis here we worked with the statistics of the last 9.5 seconds of each 30 second bout, yielding $8 \cdot 10^{4}$ states per stimulus condition, for $N=140$ cells.

\subsection*{Maximum Entropy Model Inference} 

Our maximum entropy model inference process implements a modified form of sequential coordinate gradient descent, described in \cite{dudik, broderick, tkacik2014}, which uses an L1 regularization cost on parameters. For the k-pairwise model, we inferred without a regularization cost on the local fields. Further details are given in the Supplement.

To measure the heat capacity we simulated an annealing process. Initializing at high temperatures, we monte carlo sampled half a million states per temperature level (in 100 parallel runs), initializing subsequent lower temperature runs with the final states of preceeding higher temperature runs. The heat capacity at a particular temperature was then evaluated as $C=(\langle E^{2}\rangle-\langle E\rangle^{2})/T^{2}$.

\subsection*{ Network of LN neurons}

Our model LN neurons were estimated over the chromatic checkerboard recording in the {\it light} condition. For each cell, the three color-dependent linear filters (the full STA) were weighted equally before convolution with the stimulus for an estimate of the linear response {\it q}. The non-linearity was estimated over the same data by Bayes rule, $P( \mathrm{spike} | q) = P( q | \mathrm{spike} ) P(\mathrm{spike}) / P(q)$. Spike trains were simulated from a novel pseudorandom sequence put through the model's filters and non-linearity, with the non-linearity shifted horizontally to constrain the firing rates of the neurons to be the same as in the experimental recordings. The result was binned and binarized to yield  $N=1.45 \cdot 10^5$ states, for which we inferred the k-pairwise model.

\section*{Acknowledgements}
This work was supported with a grant R01-EY014196 from the National Eye Institute to MJB and another grant 1504977 from the National Science Foundation to MJB. We thank Bill Bialek for many insightful discussions that helped determine the final form of our study. We thank Sebastiaan van Opheusden, Cengiz Pehlevan, Dietmar Plenz and Jakob Macke for helpful discussions. We also thank Eric Chen, Nikhil Deshmukh and Stephanie Palmer for help with preliminary experiments and acquisition of natural movie clips.

\end{document}


\section{Supplementary Information}

\subsection{Estimating Chromatic Sensitivities}

We estimated color-dependent spike triggered averages (STAs) for each ganglion cell. To identify the Receptive Field (RF) centers within each STA, we took the normalized dot product of each checker's time course with the time course of the extremal checker. Checkers were identified as belonging to the RF center if this dot product exceeded 0.3, and they were connected by 4-neighbor connectivity to the extremal checker. The time course of the sum over the RF centers was defined as the temporal time course of the STA.

To estimate each ganglion cell's chromatic sensitivity we found (across all three color-dependent temporal time courses) the absolute maximum deviation from zero. We then took the three values of the color dependent STA's at this time point, and normalized, to generate a 3 x 1 unit vector $\Phi_i^{(d/l)}$ representing the color sensitivity of each ganglion cell $i$ in each light condition ($d/l$). To model the chromatic sensitivity of the `red rod' or `green cone' photopigment in our experimental setup, we measured the spectral output of the monitor's red, green and blue guns directly [Ocean Optics USB 2000]. We took estimates of the red rod and green cone absorption spectrum from the literature  [Perry et al., 1991; Baylor et al., 1987; Cornwall et al., 1984], and took the dot product of these spectra with the measured monitor outputs. Our estimates of the green cone and red rod photopigment sensitivites to the red, green and blue monitor guns were $ X_g = [1.15, \, 1.23, \, 0.42] $ and $ X_r = [ 0.20, \, 1.49, \, 0.86 ]$. For example, for cell $i$ in the {\it dark} adapted condition ($d$) the projection of the STA onto the rod ($r$) was defined as $z_{i,r}^d =  X_r \cdot \Phi_i^d   / 3$. The rod-cone separation for each projection was defined as the distance of the projection to the unity line (Fig. 1B in the main text) normalized by the distance of the rod-cone overlap (red square, Fig. 1B in the main text) to the unity line.

\subsubsection{Model Inference and Fitting the Expectation Values}

We binned each ganglion cell's spike train in 20 ms time windows, assigning 0 for no spikes and 1 for one or more spikes, $r_i = [0,1]$. We denote a particular population activity pattern as $R =\{ r_i \}$. Maximum entropy models describing the distributions of time-averaged variables are fit with unitless `energies':

\begin{equation}
E_{{\mathrm pairwise}}(R) =  \sum_i^N h_i r_i + \sum_{i,j \neq i}^N J_{ij} r_i r_j
\end{equation}

\begin{equation}
E_{{\mathrm k-pairwise}}(R)=  \sum_i^N h_i r_i + \sum_{i,j \neq i}^N J_{ij} r_i r_j + \sum_{k=1}^K \delta \left (  \sum_i r_i - k \right ) \lambda_k
\end{equation}

With $\delta$ the Kronecker delta. This model is constrained to match the expectation values $\langle r_{i}\rangle$, $\langle r_{i}r_{j}\rangle$ and $\langle\sum_{i}r_{i}\rangle$, averaged over the training data.

The maximum entropy model inference process implements sequential coordinate gradient descent, described in  [Dudik et al., 2004; Broderick et al., 2007; Tka\v{c}ik et al., 2014] , which uses an L1 regularization cost on parameters, derived from the expected measurement error on the corresponding expectation values. For the k-pairwise model, we inferred without a regularization cost on the local fields. This emphasises solutions to the inference of an `independent' nature, so that terms in the interaction matrix, and the k-potential, are inferred only when necessary. 

During inference, we randomly selected four fifths of the data states for training, and withheld the rest for testing. All models presented in our work were inferred with the same preset hyperparameters:

\begin{tabular}{ c c c c c c c}
Round $\#$  & 1 & 2  & 3 & 4 & 5 & 6 \\ 
$\#$ of Iterations & 200 & 200 & 200 & 300 & 300 & 800 \\  
$\#$ of Hist MC Resamples & 7 & 5 & 3 & 1 & 0 & 0 \\
$\#$ of MC Samples (x $10^6$) & 0.2 & 0.4 & 0.65 & 0.8 & 0.9 & 1 \\ 
\end{tabular}

For example, Round 2 was preset to 200 iterations. On each iteration, $4 \cdot 10^5$ monte carlo samples were drawn to estimate the expectation values. These same samples were re-used with histogram MC resampling  [Broderick et al., 2007] 5 additional times. We converged to these choices of hyperparameters after finding that they were sufficient for the two main datasets.

On each round we also estimated the rms of all mistakes in expectation values (normalized by the appropriate error estimate, i.e. these were z-scores) that we were fitting, excluding those rare expectation values which were zero in the training data. If the rms of z-scores fell below 1 on a particular round, that round exited early, and we progressed to the next set of hyperparameters. These early exits were implemented to avoid overfitting in the networks with artifically low correlation (such as the $\alpha = 0.1$ network in the scaled covariance manipulation in the main text).

\begin{figure}
\includegraphics[width=8.7cm]{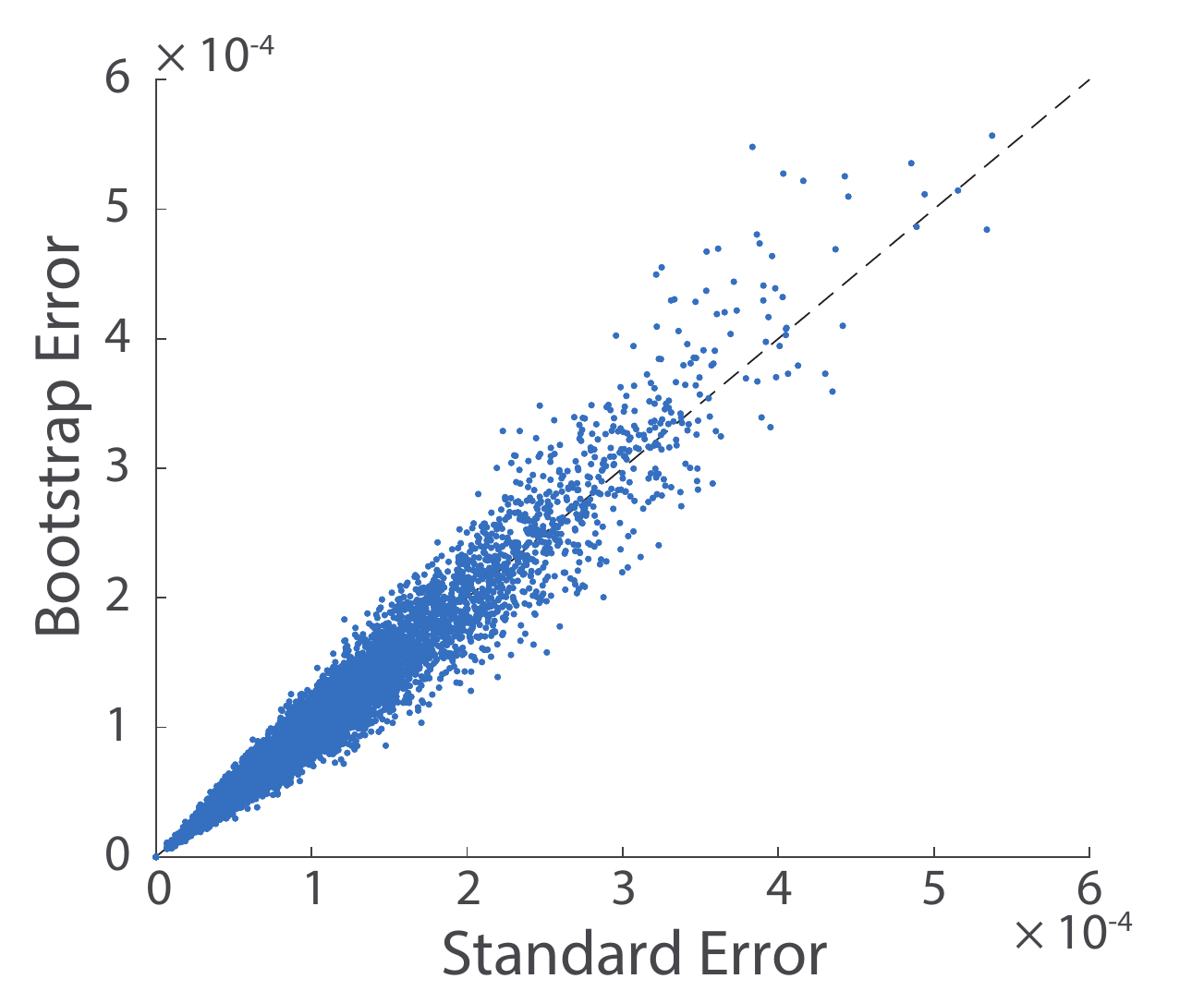}
\begin{caption}{{\bf Comparing Error Estimates}. Error estimate for the pairwise expectation value $\langle r_i r_j \rangle$ in the {\it light} dataset, for each pair of cells, over $N=128 \cdot 127/2$ pairs. Bootstrap resamples were estimated as standard deviations over 30 resamples of the data. The standard error was given by the standard deviation over samples, divided by the square root of the number of samples ($N=1.4 \cdot 10^5$ training data states). The dashed line is the unity line.}
\end{caption}
\label{fig:sfig1}
\end{figure}

The error estimate in the z-scores had contributions from both error estimates in the testing data and MC samples drawn from the model. We used the standard error of the mean as estimates of the error, because bootstrap resampling to estimate the errors of our expectation estimates was not always an available option. For example, in the manipulation where we scaled the covariance matrix by a constant $\alpha$, we had direct access to the new covariance matrix, but there were no `data' states to resample. Thus for consistency across all inference procedures we worked with the standard error of the mean. We have found the two estimates of variability to be comparable in our data (Fig. S1). 

We found that fit quality improvement (estimated on the testing data) slowed substantially after the first several thousand parameter updates [Fig. S2 E,G and Fig. S3 E,G]. Note that the error bar depends on the number of samples in both the data (there was 4X more training data than testing data), and in the model (the number of MC samples on a given iteration). Thus on each subsequent increase in MC sample number the estimated errors decreased and the z-scores increased.

In detail, we show the quality of fits for the two datasets. These are from the recording in Experiment $\#1$, for the natural movie stimulus (M1), in the two light conditions (Fig. S2 and Fig. S3, panels A-G).

\begin{figure}
\includegraphics[width=17.8cm]{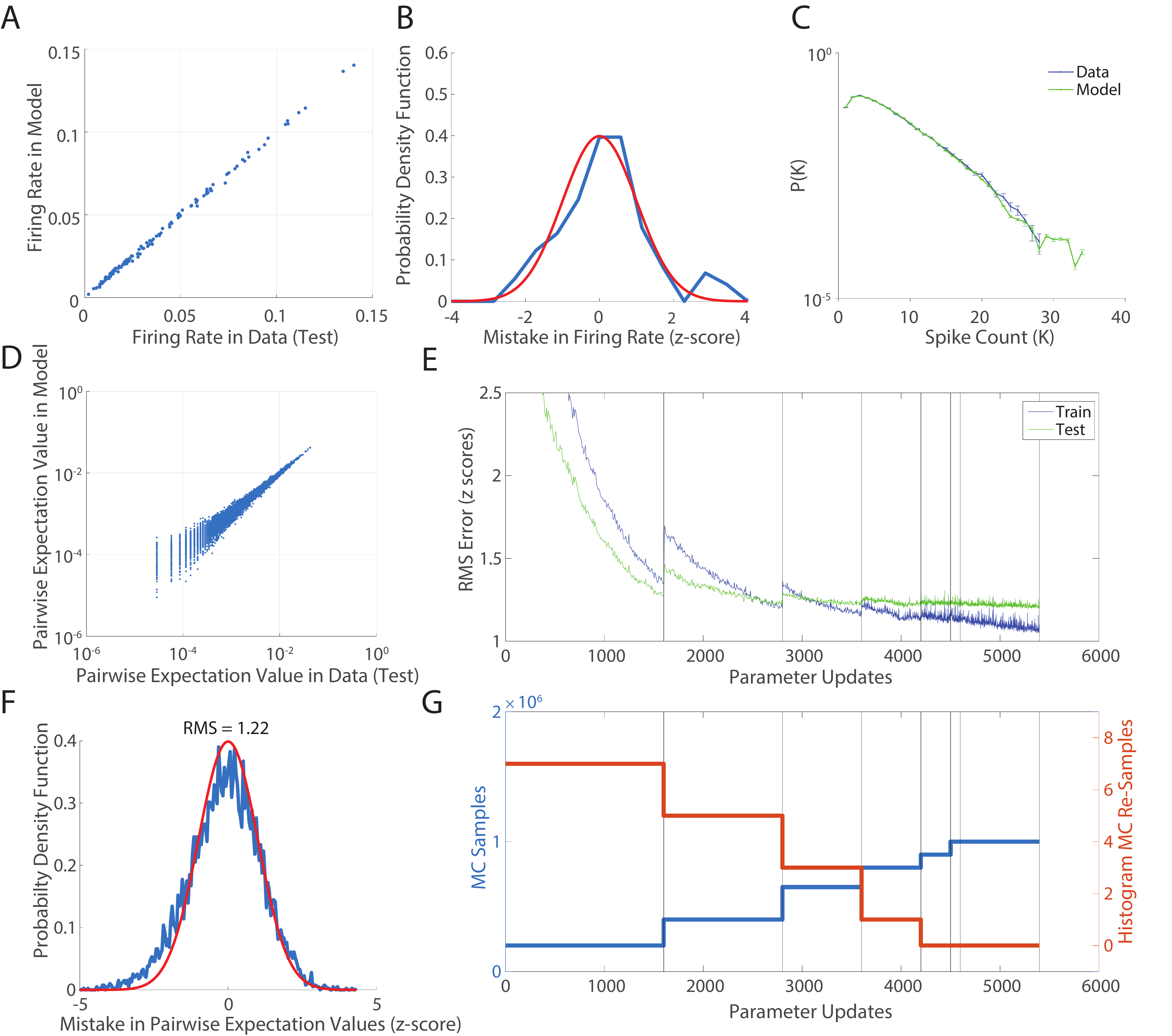}
\begin{caption}{{\bf Quality of Fitting. Experiment $\#1$, M1, Light.} $N=128$ ganglion cells and a total of $N = 1.75 \cdot 10^5$ states.  The estimates of statistics in the data here use withheld (testing) data. {\bf A}. Matching the firing rates estimated in the data with the model. These are expectation values over the binarized 20ms bins. Error bars are not shown, see panel ({\bf B}). {\bf B}. Distributions of the z-scores of mistakes in estimating the firing rates in panel {\bf A}, with a numerical gaussian in red. {\bf C}. Matching P({\it K}) in the model (green) to that estimated from the data (blue). Error bars are standard error of the mean. {\bf D}. Pairwise expectation values ($\langle r_i r_j \rangle$) estimated from the model (y-axis) plotted against the pairwise expectation values measured experimentally (x-axis), for all pairs of cells. Note the log scale. No error bars plotted, but see ({\bf F}). {\bf E}. The root mean square (RMS) of the z-scores (estimated over all expectation values that were non zero in the data) during inference. {\bf F}. The distribution of the z-scores (in blue), and in red a numerical gaussian for comparison. The RMS of these z-scores is $1.22$. {\bf G}. The number of MC samples and histogram resamples on each round during {\bf E}.}
\end{caption}
\label{fig:sfig2}
\end{figure}

\begin{figure}
\includegraphics[width=17.8cm]{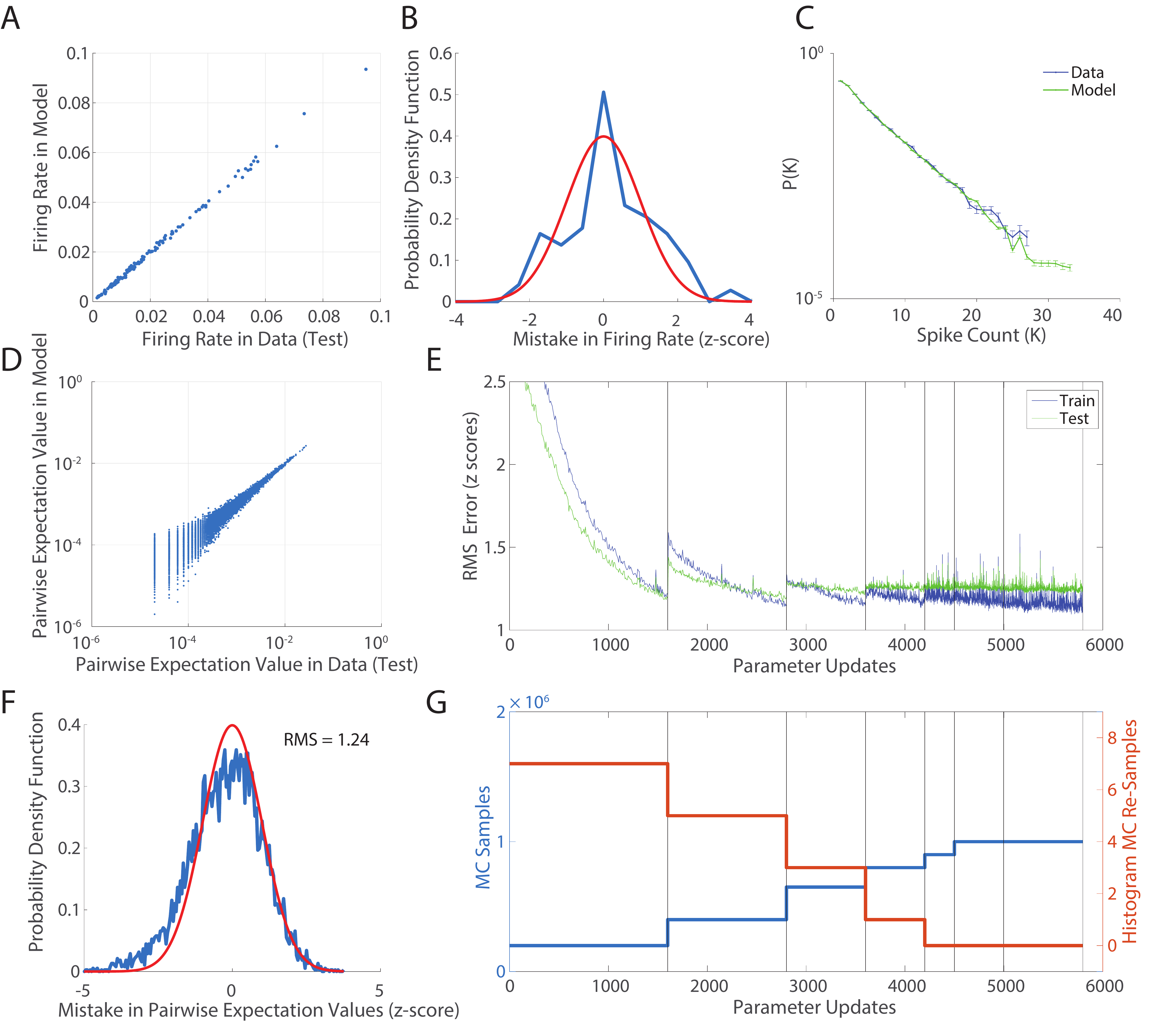}
\begin{caption}{{\bf Quality of Fitting. Experiment $\#1$, M1, Dark.} $N=128$ ganglion cells and a total of $N = 2.5 \cdot 10^5$ states. The estimates of statistics in the data here use withheld (testing) data. {\bf A}. Matching the firing rates estimated in the data with the model. These are expectation values over the binarized 20ms bins. Error bars are not shown, see panel ({\bf B}). {\bf B}. Distributions of the z-scores of mistakes in estimating the firing rates in panel {\bf A}, with a numerical gaussian in red. {\bf C}.  Matching P({\it K}) in the model (green) to that estimated from the data (blue). Error bars are standard error of the mean. {\bf D}. Pairwise expectation values ($\langle r_i r_j \rangle$) estimated from the model (y-axis) plotted against the pairwise expectation values measured experimentally (x-axis), for all pairs of cells. Note the log scale. No error bars plotted, but see ({\bf F}).  {\bf E}. The root mean square (RMS) of the z-scores (estimated over all expectation values that were non zero in the data) during inference. {\bf F}. The distribution of the z-scores (in blue), and in red a numerical gaussian for comparison. The RMS of these z-scores is $1.24$. {\bf G}. The number of MC samples and histogram resamples on each round during {\bf E}.}
\end{caption}
\label{fig:sfig3}
\end{figure}

\subsubsection{Matching Other Statistics In the Distributions of Neural Responses}

To test how well our models captured aspects of the distribution that were not explicitly fit by the models, we followed two recent publications  [Tka\v{c}ik et al., 2014; Ganmor et al., 2011]. 

We estimated the triplet correlations $C_{ijk} = \langle (r_i - \langle r_i \rangle )(r_j - \langle r_j \rangle )(r_k - \langle r_k \rangle )\rangle$ over all states in the data, and over $5 \cdot 10^5$ MC samples drawn from the model. These were then sorted by value in the data, and grouped into one thousand equally populated groupings. The mean and standard deviation over those groupings were close to the unity line (Panel A in Fig. S4,5). These results appear very similar to those in [Tka\v{c}ik et al., 2014] (there is a factor of 8 shift in the absolute values of $C_{ijk}$ arising from a change in variables from their spin notation, [-1,+1], to our spiking notation, [0,1]).

Maximum entropy models also do fairly well in matching the conditional probability of a spike  [Tka\v{c}ik et al., 2014]. Specifically, for any given state $R$, each cell $i$ has a predicted probability of a spike: 

\begin{equation}
x_i (R) = \mathrm{exp}(h_i^{\mathrm{eff}}  (R)) / ( \mathrm{exp}(h_i^{\mathrm{eff}} (R)) +1)
\end{equation}

\begin{equation}
h_i^{\mathrm{eff}}  (R) = h_i + \sum_{j \neq i} J_{ij} r_j + \lambda_{K+1} - \lambda_K
\end{equation}

( $K = \sum_{j \neq i} r_j$ ). 

Following [Tka\v{c}ik et al., 2014] we compared how well the model captured the experimental conditional probabilities (Panel B in Fig. S4,5). To do this, we took all the states in the data, and estimated the effective field for every state for every cell. We then sorted the actual binary activities by the strength of the effective field, and averaged these populations of activities in equally spaced bins (black dots). Error bars are standard deviations over the appropriate populations. These should be compared to the model prediction, which is in red (a logistic function). The gray shaded area is the probability density function of the effective field values. Inset is the same graph, on a log scale, to demonstrate that for low effective fields the conditional probabilities are well matched.

We also estimated the ability of our models to capture the probabilities of particular states. Here we compared our results to [Ganmor et al., 2011] (Panel C in Fig. S4,5). All states in the data were grouped by their unique probability (estimated directly from the data). Over these groups, we estimated the average and standard deviation of the log likelihood ratio of the model to the data. The shaded area corresponds to the $95\%$ confidence interval given by a binary distribution.

Our model fits performed well in capturing these higher order statistics, with comparable quality to the fits in previously published results [Tka\v{c}ik et al., 2014; Ganmor et al., 2011].

\begin{figure}
\includegraphics[width=17.8cm]{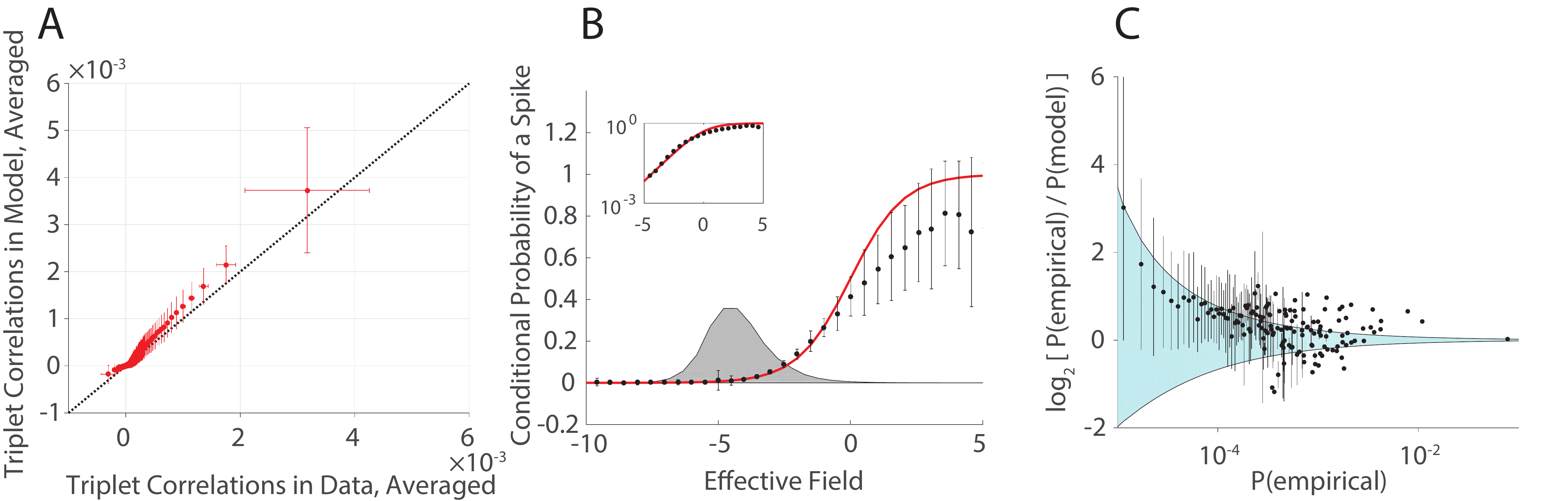}
\begin{caption}{{\bf Matching Other Statistics in the Distribution.} Experiment $\#1$, M1, Light. $N=128$ ganglion cells and $N= 1.75 \cdot 10^5$ states. {\bf A,B} follow  [Tka\v{c}ik et al., 2014]. {\bf A}. Comparison of averaged triplet correlations $C_{ijk} = \langle (r_i - \langle r_i \rangle )(r_j - \langle r_j \rangle )(r_k - \langle r_k \rangle )\rangle$ in the data and the model. {\bf B}. Comparison of model prediction of activity (based on effective fields) to the actual occurences of spikes in the data.  {\bf C}. Following [Ganmor et al., 2011] we estimated how well the model captures the probabilities of states occuring in the data (empirical). The shaded area corresponds to the $95\%$ confidence interval given by a binary distribution. }
\end{caption}
\label{fig:sfig4}
\end{figure}

\begin{figure}
\includegraphics[width=17.8cm]{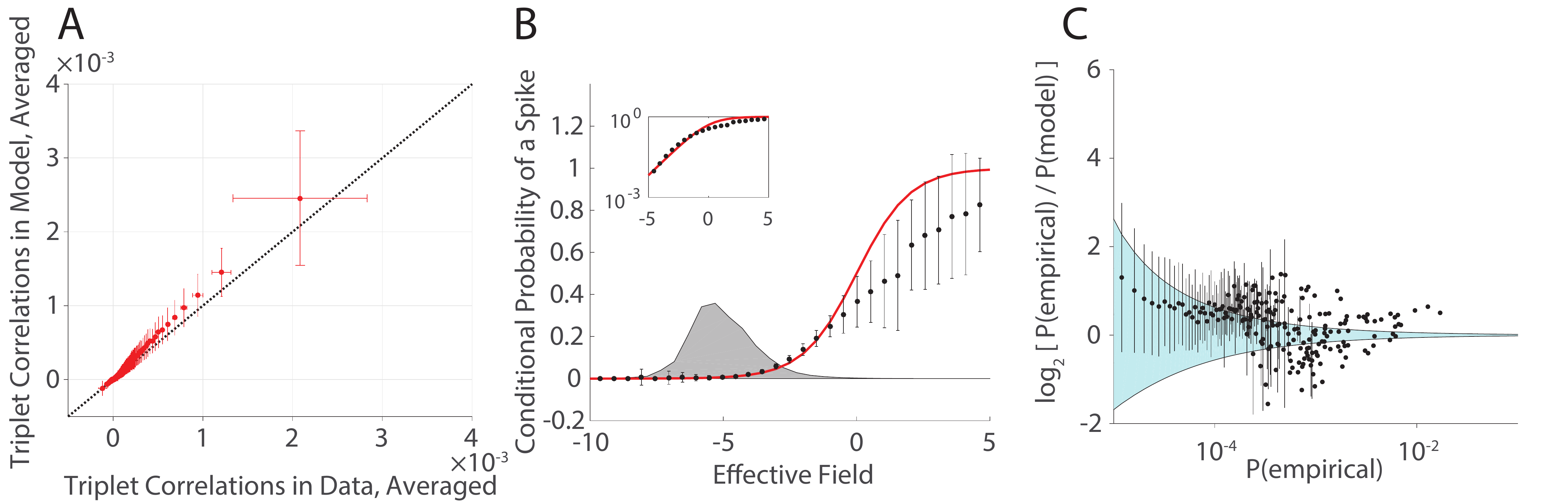}
\begin{caption}{{\bf Matching Other Statistics in the Distribution.} Experiment $\#1$, M1, Dark. $N=128$ ganglion cells and $N= 2.5 \cdot 10^5$ states. {\bf A,B} follow  [Tka\v{c}ik et al., 2014]. {\bf A}. Comparison of averaged triplet correlations $C_{ijk} = \langle (r_i - \langle r_i \rangle )(r_j - \langle r_j \rangle )(r_k - \langle r_k \rangle )\rangle$ in the data and the model. {\bf B}. Comparison of model prediction of activity (based on effective fields) to the actual occurences of spikes in the data. {\bf C}. Following [Ganmor et al., 2011] we estimated how well the model captures the probabilities of states occuring in the data (empirical). The shaded area corresponds to the $95\%$ confidence interval given by a binary distribution. }
\end{caption}
\label{fig:sfig5}
\end{figure}

\subsubsection{Analytical vs. Model Fits for Independent Networks}

The analytical solution for independent neurons consists of a local field to each neuron ($h_i^{\mathrm{ind}} = -\log ( \langle r_i \rangle / (1-\langle r_i \rangle ) )$), which depends solely on the measured firing rates ($\langle r_i \rangle$). All other terms ($J_{ij}$, $\lambda_k$) are zero. However, for shuffled data, inference returned models with non-zero entries in the interaction matrix, and k-potential.  Because of this discrepancy we compared the shuffled data model to the analytical solution by comparing the expectation values and shapes of the specific heat predicted by both (Fig. S6).

\begin{figure}
\includegraphics[width=17.8cm]{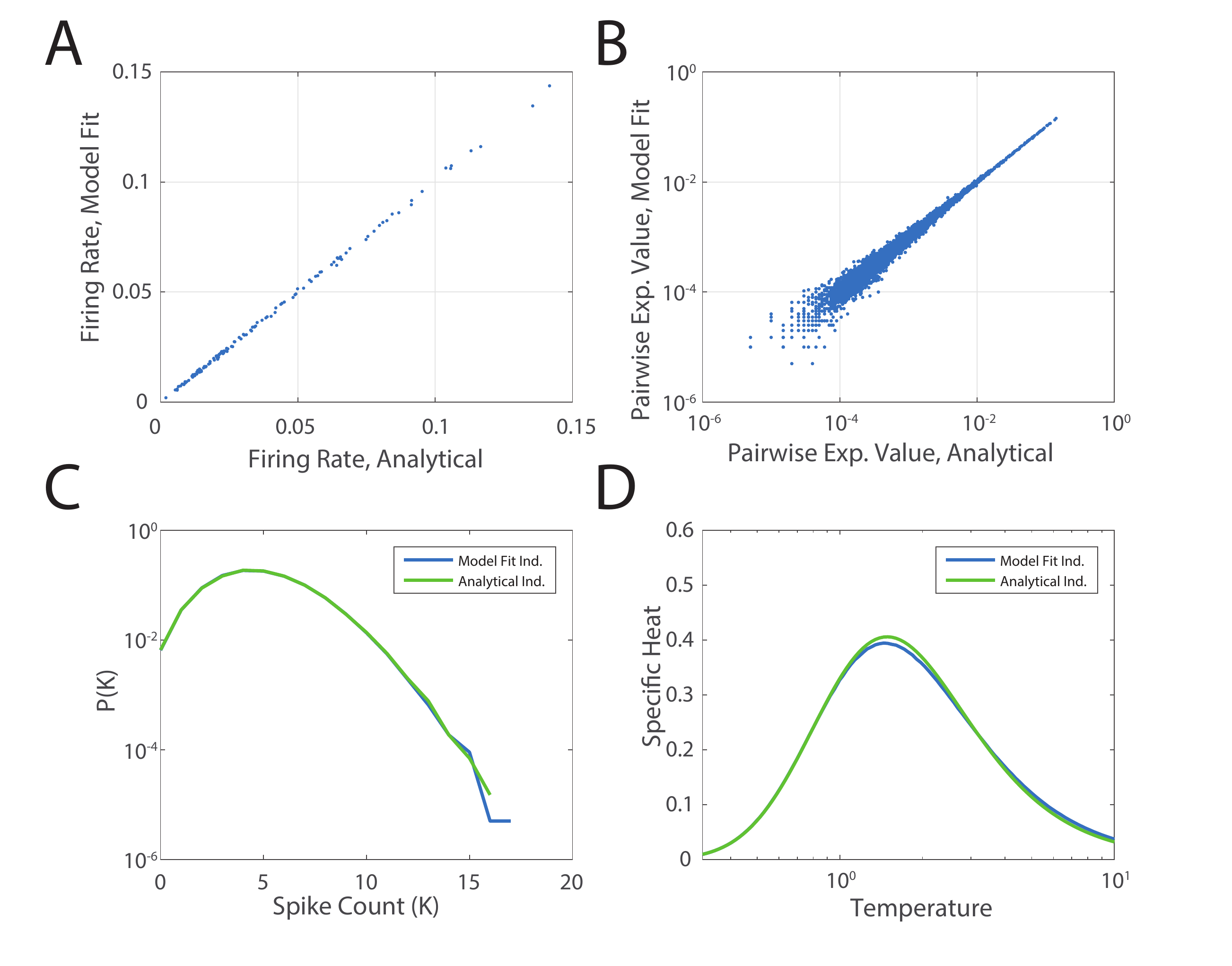}
\begin{caption}{{\bf Comparing Analytical Solutions to Shuffled-Data Model Fits}. For the analytical solution, we took all the states in the {\it light} condition ($N=1.75 \cdot 10^5$), to estimate the average activity of each neuron. For the shuffled data, model parameters were inferred on a training subset of all of those spikes, as described earlier in the Supplement. {\bf A.} Firing rates (estimated over 20ms time bins) compared across the two models, for $N=128$ cells in the data. {\bf B.} Pairwise expectation values ($\langle r_i r_j \rangle $, estimated over 20ms binarized bins) compared across the two models. {\bf C.} Distribution of Spike Counts, $P(K)$, for the two models. {\bf D.} Specific heat as a function of temperature, for the two models. }
\end{caption}
\label{fig:sfig6}
\end{figure}

\subsubsection{Pairwise Maximum Entropy Model Results}

Our main result concerning the presence of a phase transition was qualitatively similar in pairwise maximum entropy model fits to the data. As in the main text, the pairwise maximum entropy model was fit to subsets of cells in the {\it light} and {\it dark} datasets. A fictitious temperature was introduced, and the variance of energy levels was evaluated over Monte Carlo samples from these distributions. The specific heat at different system sizes for the {\it light} dataset is plotted in Fig. S7A, for the {\it dark} in Fig. S7B. The peak temperature, $T_{max}$, and peak value, $C(T_{max})$, of the specific heat are plotted as a function of the inverse size of the analyzed neural population, $1/N$ (Fig. S7C,D). As for the k-pairwise model, the peak sharpens, yet is always to the right of the operating point of the network, suggesting that the network is on the low-temperature side of the phase transition.

\begin{figure}
\includegraphics[width=17.8cm]{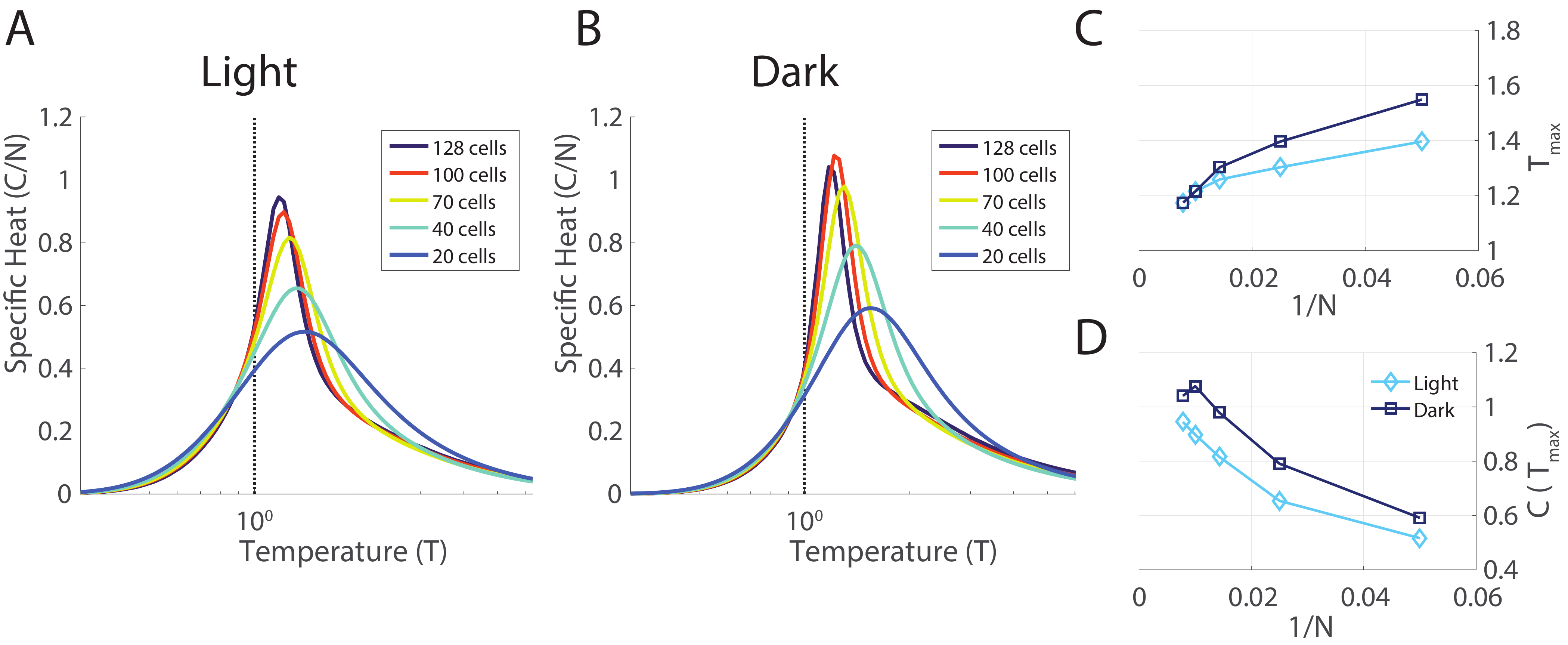}
\begin{caption}{{\bf Pairwise Maximum Entropy Model Results for Natural Movies} (Experiment $\#1$, M1, {\it light} and {\it dark}, $N=128$ ganglion cells). All manipulations are identical to those done in the main text. {\bf A}. The specific heat for subsampled populations in the {\it light} condition. {\bf B}. The specific heat for subsampled populations in the {\it dark} condition. {\bf C,D}. The peak temperature, $T_{max}$, and peak value, $C(T_{max})$, of the specific heat are plotted as a function of the inverse size of the analyzed neural population, $1/N$. }
\end{caption}
\label{fig:sfig7}
\end{figure}

\subsubsection{Persistence Indices}

We saw in the main text that the persistence index for any given neural activity state correlated well with its dwell time estimated by $T=1$ MC sampling dynamics (Fig. 9C).

We show here a plot of the persistence indices from the same neural activity state measured in the light vs dark conditions, which exhibited high correlation (Fig. S8A). This correlation is evidence that the energy landscape has a similar shape in light and dark conditions, even though the specific details of the neural code change considerably (Fig. 2). This similarity is important, because the stimulus was the same in both conditions. We expect that the retina should encode many of the same stimulus features across different light levels. One possibility is that the retina uses different codewords across different light levels, in which case the brain must change its decoding strategy as a function of the light level. The other possibility is that some aspects of the retinal code remains invariant across light levels.  While no such invariance was obvious at the level of single neurons (see Fig. 1 and [Tikidji-Hamburyan et al., 2015]), these data provide the first evidence that a form of light-level invariance does exist at the population level.

For comparison, we also plot the persistence indices estimated from independent models in the {\it light} and {\it dark} adapted conditions (Fig. S8B). Here, the correspondence between the neural code in light and dark conditions was poor, with a correlation coefficient = 0.37 in Fig. S8B vs 0.90 in Fig. S8A).

Clearly, the form of invariance to light level that we observed in the population neural code (Fig. S8A) is a consequence of the correlations in the experiment. To test the reproducibility of this measure, we split the {\it light} data into two random, complementary halves, and inferred the k-pairwise maximum entropy model parameters for these halves two separately. The persistence indices display a high degree of similarity across these two random halves (Fig. S8C), indicating that this a highly reproducible measure that can be well sampled in our data.  The comparison with Fig. S8A also shows that the invariance of the population code to light level nearly approaches the limit set by sampling noise.

\begin{figure}
\includegraphics[width= \textwidth]{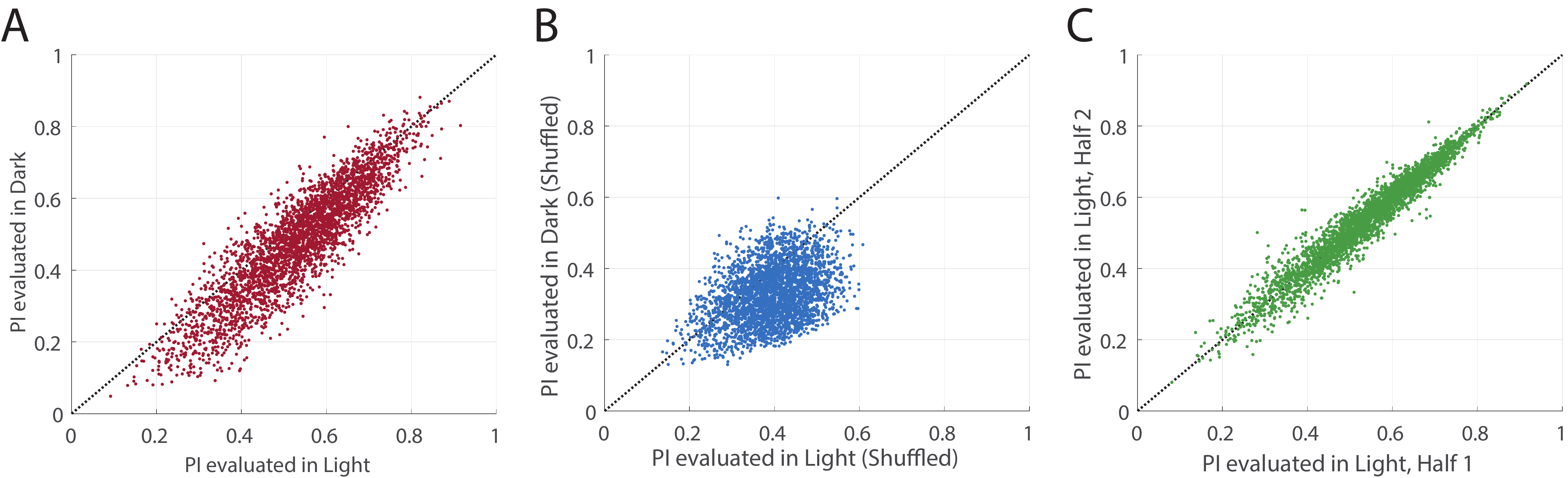}
\begin{caption}{{\bf Persistence Indices}. Here, we compare the persistince indices for the $N=3187$ states from the {\it light} natural movie recording with $K=12$, using different model parameters. {\bf A}. PI estimated by the full model fitting the natural movie in the {\it light} (x-axis) and {\it dark} (y-axis) adapted conditions (correlation coefficient of 0.90). {\bf B}.  PI estimated by the independent model (shuffled data) fitting the natural movie in the {\it light} (x-axis) and {\it dark} (y-axis) adapted conditions (correlation coefficient of 0.37). {\bf C}. PI estimated from model fits to two complementary random halves of the {\it light} dataset (correlation coefficient of 0.97). }
\end{caption}
\label{fig:sfig8}
\end{figure}

\subsubsection{Zipf-like Relationships in the Distribution of Codewords}

A separate line of work has investigated the presence of Zipf-like relationships in the probability distribution of neural codewords  [Mora \& Bialek, 2011; Schwab et al., 2014]. We largely avoided working with these quantities out of concerns about the adequacy of sampling the we could achieve. In Fig. S9 we plot the Zipf relationship between the probability and rank of a state (estimated directly from the natural movie dataset in Experiment $\#1$). While the slopes of a power law fit to our experimental data were close to -1, it is important to note that a clear power law relationship existed only over a small range of rank $\sim 200$ to 1000. This leaves a substantial uncertainty in the estimate of the slope. But more fundamentally, we are not sure how to interpret deviations from a slope of -1 as well as how to think about deviations from the power law form itself, such as the "bump" near rank $\sim 70$ in the dark and near rank $\sim 300$ in the light (Fig. S9).

\begin{figure}
\includegraphics[width=17.8cm]{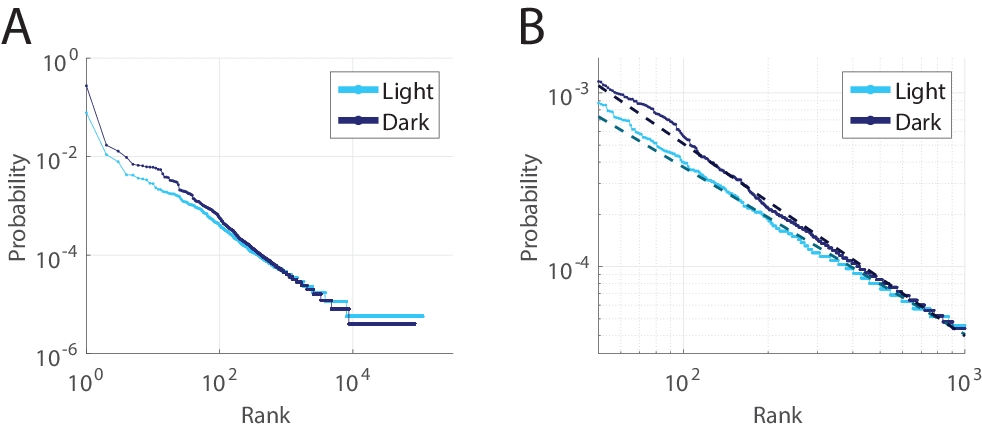}
\begin{caption}{{\bf Zipf Relationships}. Probability vs. Rank for States from the {\it light} (light blue) and {\it dark} (dark blue) datasets evaluated directly. On the left, all states are plotted for both datasets. On the right, the probabilities of the states ranked from 50 to 1000 only are plotted, and a linear fit to this subset of states is shown in dashed lines. The fitted slopes are -0.96 for the {\it light} and -1.11 for the {\it dark} datasets .}
\end{caption}
\label{fig:sfig9}
\end{figure}

\pagebreak
\subsection{References}

Baylor, D. A., Nunn, B. J., \& Schnapf, J. L. (1987). Spectral sensitivity of cones of the monkey Macaca fascicularis. {\it The Journal of Physiology, 390}, 145.

Broderick, T., Dudik, M., Tkacik, G., Schapire, R. E., \& Bialek, W. (2007). Faster solutions of the inverse pairwise Ising problem. {\it arXiv preprint arXiv:0712.2437}.

Cornwall, M. C., MacNichol, E. F., \& Fein, A. (1984). Absorptance and spectral sensitivity measurements of rod photoreceptors of the tiger salamander, Ambystoma tigrinum. {\it Vision research, 24}(11), 1651-1659.

Dudik, M., Phillips, S. J., \& Schapire, R. E. (2004). Performance guarantees for regularized maximum entropy density estimation. In {\it Learning Theory} (pp. 472-486). Springer Berlin Heidelberg.

Ganmor, E., Segev, R., \& Schneidman, E. (2011). Sparse low-order interaction network underlies a highly correlated and learnable neural population code. {\it Proceedings of the National Academy of Sciences, 108}(23), 9679-9684.

Macke, J. H., Opper, M., \& Bethge, M. (2011). Common input explains higher-order correlations and entropy in a simple model of neural population activity. {\it Physical Review Letters, 106}(20), 208102.

Mora, T., \& Bialek, W. (2011). Are biological systems poised at criticality?. {\it Journal of Statistical Physics, 144}(2), 268-302.

Perry, R. J., \& McNaughton, P. A. (1991). Response properties of cones from the retina of the tiger salamander. {\it The Journal of Physiology, 433}, 561.

Schwab, D. J., Nemenman, I., \& Mehta, P. (2014). Zipf’s law and criticality in multivariate data without fine-tuning. {\it Physical review letters, 113}(6), 068102.

Sherry, D. M., Bui, D. D., \& Dgerip, W. J. (1998). Identification and distribution of photoreceptor subtypes in the neotenic tiger salamander retina. {\it Visual neuroscience, 15}(06), 1175-1187.

Tikidji-Hamburyan, A., et al. (2015). Retinal output changes qualitatively with every change in ambient illuminance. {\it Nature neuroscience, 18} (1), 66-74.

Tka\v{c}ik, G., et al. (2014). Searching for collective behavior in a large network of sensory neurons.  {\it  PLoS Comput Biol, 10}(1), e1003408.